\documentclass[aps,twocolumn,pra,groupedaddress,superscriptaddress,nofootinbib]{revtex4-2}

\usepackage{graphicx,amsmath,amssymb,amsbsy,subfigure,hyperref,bbm,times,txfonts,float}
\usepackage{subfigure,hyperref,bbm,times}
\usepackage[T1]{fontenc}
\usepackage{braket}
\usepackage{epsfig} 
\usepackage{longtable}
\usepackage{cleveref}
\usepackage{color}
\usepackage{graphicx}
\usepackage{dcolumn}
\usepackage{bm}

\providecommand{\openone}{\leavevmode\hbox{\small1\kern-3.8pt\normalsize1}}

\usepackage{soul}
\usepackage{cancel}
\usepackage{float}
\usepackage{multirow}
\newcolumntype{?}{!{\vrule width 1.8pt}}

\usepackage{colortbl}

\hypersetup{
	colorlinks=true,
	linkcolor=blue,
}
\graphicspath{{figure/}}

\begin{document}

\title{Generation of genuine multipartite entangled states via indistinguishability of identical particles}

\author{Kobra Mahdavipour}
\email{kobra.mahdavipour@inrs.ca}
\affiliation{Dipartimento di Ingegneria, Università degli Studi di Palermo, Viale delle Scienze, 90128 Palermo, Italy}
\affiliation{Institut national de la recherche scientifique -Centre Énergie, Matériaux, et Télécommunciations (INRS-EMT), 1650 Boulevard Lionel-Boulet, Varennes, Qu\'{e}bec J3X 1S2, Canada}

\author{Farzam Nosrati}
\email{farzam.nosrati@unipa.it}
\affiliation{Dipartimento di Ingegneria, Università degli Studi di Palermo, Viale delle Scienze, 90128 Palermo, Italy}
\affiliation{Institut national de la recherche scientifique -Centre Énergie, Matériaux, et Télécommunciations (INRS-EMT), 1650 Boulevard Lionel-Boulet, Varennes, Qu\'{e}bec J3X 1S2, Canada}
\affiliation{IMDEA Networks Institute, Madrid, Spain}

\author{Stefania Sciara}
\affiliation{Institut national de la recherche scientifique -Centre Énergie, Matériaux, et Télécommunciations (INRS-EMT), 1650 Boulevard Lionel-Boulet, Varennes, Qu\'{e}bec J3X 1S2, Canada}

\author{Roberto Morandotti}
\affiliation{Institut national de la recherche scientifique -Centre Énergie, Matériaux, et Télécommunciations (INRS-EMT), 1650 Boulevard Lionel-Boulet, Varennes, Qu\'{e}bec J3X 1S2, Canada}

\author{Rosario Lo Franco}
\email{rosario.lofranco@unipa.it}
\affiliation{Dipartimento di Ingegneria, Università degli Studi di Palermo, Viale delle Scienze, 90128 Palermo, Italy}

\begin{abstract}

Indistinguishability of identical particles is a resource for quantum information processing and has been utilized to generate entanglement from independent particles that spatially overlap only at the detection stage. {Here, we introduce a controllable scheme capable of generating via three steps, initialization, deformation, and post-selection, different classes of multipartite entangled states starting from a product state of $N$ spatially distinguishable identical qubits. While our scheme is generalizable to any class of entangled bosonic and fermionic systems, we provide an explicit recipe for the generation of W, Dicke, GHZ, and cluster states, which are resource states for quantum information processing. Using graph-based representations within the framework of spatially localized operations and classical communication (sLOCC), we mathematically demonstrate a direct translation of the generation schemes of specific entangled states into colored, complex, and weighted digraphs, each corresponding to a given experimental setup. We also show that this graph-theoretical approach allows for the optimization of the generation efficiency of specific multipartite entangled states by exploring a variety of generation schemes. The presented theoretical approach, while already implementable with current linear optics architectures, has the potential to bring clear advantages over existing technologies, such as in quantum computing search algorithms and in the design of new experiments in quantum optics or other platforms. }

\end{abstract}

\maketitle
\section{Introduction\label{introduction}}

The outperformance of quantum technologies compared with classical ones resides in exploiting quantum effects, such as entanglement, for various applications, including computing \cite{Shor1994Algorithms, Arute2019, Zhong2020Quantum}, secure communication \cite{ekert1992quantum, Nadlinger2022Experimental}, and sensing \cite{giovannetti2011advances}. In addition, the indistinguishability associated with systems of identical quantum entities (e.g., photons, electrons, and atoms of the same species) is a unique notion in quantum mechanics. Particles are named indistinguishable when their wavefunctions become spatially overlapped and their global state does not allow to address them individually \cite{peres1997quantum}. Consequently, these particles behave differently and combine together to form compound structures of matter and light, in which the states of identical bosons (fermions) are symmetric (antisymmetric) under the exchange of any pair of particles \cite{peres1997quantum}. This intrinsic property is responsible for distinct phenomena such as electron orbital occupation \cite{pauli2012general} and photon bunching \cite{hong1987measurement}.

Beyond the fundamental aspects, indistinguishability can be a key resource for quantum-enhanced applications. For instance, spatially localized operations and classical communication (sLOCC) \cite{franco2018indistinguishability} have been proposed to exploit indistinguishability for tasks such as quantum teleportation \cite{franco2018indistinguishability,sun2020experimental}, quantum metrology \cite{castellini2019indistinguishability, Sun2022Activation}, and communication in multimode photonic quantum networks \cite{bellomo2017n, Activating2019Castellini, Wang2022Remote}. Also, indistinguishability offers a shield to protect quantumness against various noise sources \cite{perez2018endurance, nosrati2020robust, nosrati2020dynamics, piccolini2021entanglement, piccolini2021indistinguishability, Piccolini2023, piccolini2023asymptotically,piccolini2024,nosrati2024QST}. An elementary entangling gate can be achieved by spatially overlapping two independent particles with opposite pseudospins that meet only at the detectors, as proven both theoretically \cite{franco2018indistinguishability} and experimentally \cite{sun2020experimental,barros2020entangling}. However, these generation schemes have been so far limited to bipartite \cite{franco2018indistinguishability,sun2020experimental,barros2020entangling} and tripartite cases \cite{Lee:22}.

Unlike bipartite systems, multipartite entangled systems exhibit much more complex correlation structures \cite{guhne2009entanglement, Horodecki2009Quantumentanglement}. Different inequivalent entanglement classes emerge in the multipartite setting, and they never convert to each other under stochastic local operations and classical communication  \cite{PhysRevA.62.062314}. Each multipartite entangled class has unique features and finds applications to specific quantum information tasks. For example, Greenberger–Horne–Zeilinger (GHZ) states \cite{PhysRevLett.115.020502} are used for ballistic quantum computing \cite{zhou2022double, PhysRevLett.115.020502} and quantum metrology \cite{Tóth_2014}, while cluster states are building blocks for various universal measurement-based quantum computers \cite{KieselExperiment2005,Raussendorf2001One,briegel2009measurement}. $W$ and Dicke states stand out because their entanglement is robust to particle loss \cite{Persistent2001Briegel, Multipartite2001Dur_W} and have applications in quantum communication, such as secret sharing \cite{KieselExperiment2007,Prevedel2009Experimental, Wieczorek2009Experimental} and quantum teleportation \cite{Joo2003Quantum}. Several platforms have been proposed to generate multipartite entangled states, such as superconducting qubits \cite{Wang2018, DiCarlo2010, PhysRevLett.119.180511, Gong2019Genuine12Qubit}, trapped ions \cite{Sackett2000, Leibfried2005,PhysRevLett.106.130506, PhysRevX.8.021012, Mandel2003, PhysRevLett.111.210501,PhysRevLett.119.150503, PRXQuantum.2.020343}, nuclear spins \cite{science.1157233}, and photonic qubits with both probabilistic \cite{Zhong2018_12_Photon, Gong2019Genuine12Qubit} and deterministic sources \cite{Schon2005Sequential, Lindner2009Proposal,Mikkel2019Deterministic, Thomas2022Efficient, science.aax9743, Besse2020}. 

{Multipartite entangled states can be generated using both deterministic and probabilistic schemes. Deterministic schemes involve either global $N$-qubit \cite{Lu2019} or $\frac{N(N-1)}{2}$ pairwise \cite{Maslov_2018} entangling gates. These schemes require precise interactions between qubits, making them unsuitable for photons, due to their weak interaction. In contrast, post-selective schemes, which rely on coherent transformations and post-selection measurements \cite{Greenberger1990, 
PhysRevLett.68.1251}, are more suitable for photonic and, more in general, remote (i.e., non-interactive) systems. Various probabilistic methods, each with different success probabilities, have been developed to generate multipartite entangled states. For example, multiple Bell pairs can be entangled to generate different classes of multipartite entangled states via exploiting indistinguishability of identical qubits, including GHZ states \cite{Greenberger1990,PhysRevLett.82.1345}. Additionally, it has been shown that the indistinguishability of identical particles, whether through spatial overlap \cite{PhysRevLett.68.1251,chin2021graph,Lee:22} or photonic path identity \cite{PhysRevLett.118.080401, RevModPhys.94.025007} formalisms, is a genuine quantum resource for generating a wide class of multipartite entangled states. Unlike photonic path identity, which relies on direct interaction (or interference) of photons at a single beam splitter and nonlinear medium \cite{RevModPhys.94.025007}, the spatial indistinguishability approach involves independent bosonic or fermionic qubits meeting (spatially overlapping) only at the detection regions \cite{franco2018indistinguishability,sun2020experimental, PhysRevApplied.18.064024}.
}

In this work, we {go beyond the case-specific designs and} provide a comprehensive theoretical method {that exploits} the indistinguishability of identical qubits {to generate different classes of} multipartite entangled states, including W, Dicke, GHZ, and cluster states. It has been shown that a graph-based representation of quantum optical experiments {\cite{QuantumExperiments2017Krenn, Quantumexperimentsandgraphs2017Xuemei, Quantumexperimentsandgraphs22017Xuemei,chin2021graph,chin2023graphs, PhysRevX.11.031044,ruiz2022digital}} can be introduced {and utilized algorithmically to facilitate the generation scheme of multipartite entangled states. A graph-theoretic approach has been used to represent the process of post-selective generation of multipartite entangled states, exploiting photons’ indistinguishability \cite{PhysRevLett.118.080401,chin2021graph}. Leveraging this connection, we establish a direct mathematical correspondence between the general generation recipe in the no-label approach and a digraph (bigraph) adjacency matrix in the graph formalism for both bosonic and fermionic particles’ statistics. This work enables the formulation of the sLOCC framework \cite{franco2018indistinguishability} in the graph-based approach, where spatial overlaps appear as adjacency matrix elements of the digraph (bigraph), to exploit the resource of indistinguishability for the generation of multipartite entanglement. This approach opens a clearer perspective on automated machine learning assistance designs \cite{Krenn2023Artificial} in generation schemes for arbitrary multipartite entangled states in various experimental platforms, such as photonic and superconducting setups, towards the optimized use of the quantum resources, including particle statistics imprints. 
}


\begin{figure*}[t!]
    \centering
\includegraphics[width=0.95\linewidth]{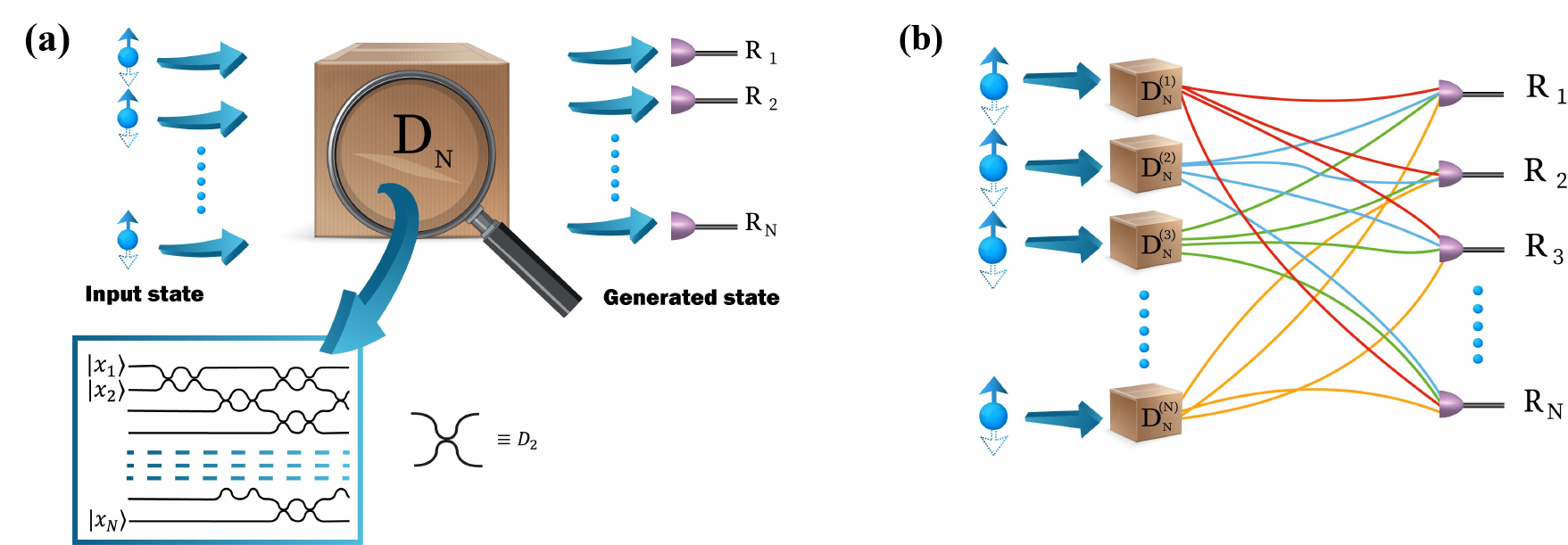}
    \caption{\textbf{Illustration of the deformation in both on-site and remote schemes.} \textbf{(a)} Illustration of an on-site designed $N$-mode deformation, denoted as $\hat{D}_{N}$, where $N$ identical qubits, each possessing an internal degree of freedom $\sigma_i=\{\uparrow,\downarrow\}$, pass through $N$ modes of this deformation. At the end, qubits arrive in $N$ detection regions $R_1, \dots, R_N$, for sLOCC projection measurement to generate the output state. The inset illustrates the potential for implementing any $N$-mode deformation by combining $D_2$ operators. \textbf{(b)}  Illustration of remote deformation design using $N$ deformations $D_N^{(i)}$ ($i=1,\ldots, N$) that send $N$-qubits into $N$ individual regions for postselection, generating the output state.}
    \label{fig: deformation illustration}
\end{figure*}

\section{Indistinguishability-based entangling gate}
The indistinguishability of identical qubits provides a fundamental quantum resource that can be utilized within the sLOCC framework to generate entanglement in a conditional fashion. In general, such a probabilistic generation scheme comprises three steps: (i) the state initialization of $N$ independent identical qubits; (ii) a general form of deformation that controllably distributes $N$ qubits into $N$ separated operational regions; (iii) a postselection process that detects one particle per region to activate the desired quantum state.

\textbf{Step I: Initialization.} The protocol starts with a $N$-particle pure product state   $\ket{\varphi_{1}}\otimes\dots\otimes\ket{\varphi_{N}}=\ket{\varphi_{1},\ldots,\varphi_{N}}$, characterized by a complete set of commuting observables of the single-qubit state $\ket{\varphi_{j}}$. Also, each single-qubit state $\ket{\varphi_{i}}=\ket{\psi_{i},\sigma_{i}}$ comprises the particle's degrees of freedom, that is, a spatial wave function $\ket{\psi_{i}}$ and a pseudospin $\ket{\sigma_{i}}=\{\uparrow,\downarrow\}$. This means that the protocol needs $N$ single identical qubits on demand.

\textbf{Step II: Deformation.} Deformations play a crucial role in generating entanglement through coherent manipulation of spatially separated particles and making them spatially indistinguishable \cite{sun2020experimental,franco2018indistinguishability}. These deformations can
be applied to both pure and mixed quantum states. Here, we are interested in the case of composite systems in pure states. One can apply a deformation on $N$ identical qubits in both on-site and remote designs, as illustrated in Fig.~\ref{fig: deformation illustration}~(a) and (b).

In the on-site design, as illustrated in Fig.~\ref{fig: deformation illustration}~(a), the $N$-particle state evolves using a single unitary deformation operator $\hat{D}_{N}$. We assume that each particle is sent through one of the input modes, denoted as $\ket{x_j} = \ket{\varphi_{j}}$, as illustrated in Fig.~\ref{fig: deformation illustration}~(a). Adopting the no-label formalism for identical particles \cite{Lo_Franco2016Quantum, Sciara2017,compagnoRSA}, the action of the on-site deformation on the N-particle
state can be written as $\hat{D}_{N}\ket{\varphi_{1},\dots,\varphi_N}=\ket{\varphi_{1D},\dots,\varphi_{ND}}=\ket{\Psi_D^{(N)}}$, where $\ket{\varphi_{iD}}$ and $\ket{\Psi_D^{(N)}}$ represent the single-qubit and $N$-qubit deformed states, respectively. The single-qubit deformed states $\ket{\varphi_{iD}}$ can be given as $\ket{\varphi_{iD}}=\sum_{j}r_{ij}\ket{R_i\:\sigma_j}$, where $r_{ij}$ are meant as the probability amplitudes that the qubit originally coming from the state $\ket{\varphi_i}$ ends up in the spatial mode $\ket{R_j}$. Notice that, since particles spatially overlap at the detection regions, the global deformed state is not generally factorizable in terms of single-qubit states $\ket{\Psi_D^{(N)}}\neq\ket{\varphi_{1D}}\otimes\dots \otimes \ket{\varphi_{ND}}$ meaning that the state vector is a whole entity \cite{compagnoRSA}. In passive deformation, which we name spatial deformation, $N$ identical qubits are spatially distributed in $N$ regions, and their pseudospin remains untouched. Differently, in active deformation, qubits are not only spatially distributed in $N$ distinct regions, but their pseudospins are also controllably altered. As sketched in Fig.~\ref{fig: deformation illustration}~(a), any $N$-mode unitary transformation $\hat{D}_N$ can be constructed using a sequence of 2-mode unitary operators $\hat{D}_2$ \cite{PhysRevLett.73.58}.  For example, one can observe the effect of a two-mode spatial deformation in an optical setup using a beam splitter.

In the remote design in Fig.~\ref{fig: deformation illustration}~(b), there are $N$ deformations $\hat{D}_N^{(i)}$ ($i=1\dots N$) that are independently applied to each qubit originally in the state $\ket{\varphi_i}$. Practically, each deformation acts on an $N$-mode input state consisting of $N-1$ vacuum mode $\ket{vac}$ and a single-qubit state $\ket{\varphi_i}$, that is $\hat{D}_{N}^{(i)}\ket{X_{N}^i}=\hat{D}_{N}^{(i)}\ket{\varphi_{i},vac,\ldots,vac}:=\hat{D}_{N}^{(i)}\ket{\varphi_{i}}=\ket{\varphi_{iD}}$, where $\ket{\varphi_{iD}}=\sum_{j=1}^{M}r_{ij}\ket{R_j^{(i)}\:\sigma_j}$ is the single-qubit deformed state with different spatial modes $R_j^{(i)}$. The overall unitary $N$-mode deformations are now written as $\hat{\mathcal{D}}_{N}^{(N)}=\hat{D}_{N}^{(1)} \otimes\cdots \otimes \hat{D}_{N}^{(N)}$. Its action on the initial pure product state
of $N$ identical qubits is $\hat{\mathcal{D}}_{N}^{(N)}\ket{\varphi_{1},\varphi_{2},\ldots,\varphi_{N}}:=\ket{\hat{D}_{N}^{(1)}\varphi_{1}\ldots,\hat{D}_{N}^{(N)}\varphi_{N}}=\ket{\varphi_{1D},\ldots,\varphi_{ND}}=\ket{\Psi_D^{(N)}}$. Note that each deformation $\hat{D}_{N}^{(i)}$ is a unitary operation, as any two single-qubit deformed states (for example, states $\ket{\varphi_i}$ and $\ket{\varphi_j}$) are orthogonal. However, since the measurement is performed in the same spatial region, it cannot resolve different spatial modes when qubits reach one detector in the same region, as illustrated in Fig.~\ref{fig: deformation illustration}~(b).  This is why, for simplicity, we will write $R_j$ instead of $R_j^{(i)}$ in the deformed states to clarify that identical qubits will be detected in the same spatial regions. As a result of the detection process in the remote design, the $N$-qubit deformed state becomes unnormalized, and we need to renormalize the state as $\ket{\Psi_D^{(N)}}\rightarrow \frac{1}{\sqrt{\nu}}\ket{\Psi_D^{(N)}}$, where the normalization factor, given as $\nu=\vert\langle \varphi_{1D},\ldots,\varphi_{ND}\vert \varphi_{1D},\ldots,\varphi_{ND}\rangle\vert^2$, takes into account the spatial overlap between single-qubit states \cite{compagnoRSA}.

As a result of this deformation, in both on-site and remote designs, identical qubits are distributed over $N$ separated measurement regions in a global deformed state $\ket{\Psi_{D}^{(N)}}$.

\textbf{Step III: Activation.} As illustrated in Fig.~\ref{fig: deformation illustration}, after the qubits have been distributed by the deformation, the sLOCC measurement process is applied to activate the desired $N$-qubit state. The measurement can be realized via single-qubit counting, which is insensitive to the internal degrees of freedom, i.e., pseudospin.

Thus, the desired $N$-qubit state is activated through the sLOCC projection \cite{nosrati2020robust}, which simultaneously performs single-qubit counting in the $N$ remote spatial regions. This projection is given by the operator
\begin{equation}\label{eq: sLOCC proj}
    \hat{\Pi}^{(N)}=\sum_{k}\ket{R_1\sigma_1^{(k)},\ldots,R_N\sigma_N^{(k)}}\bra{R_1\sigma_1^{(k)},\ldots,R_N\sigma_N^{(k)}},
\end{equation}
where the index $k:=\{\sigma_1^{(k)},\ldots,\sigma_N^{(k)};\ \sigma_j^{(k)}= \uparrow, \downarrow\}$ runs over all the $2^N$ pseudospin combinations. Necessarily, the measurement device cannot access the other degrees of freedom (e.g., pseudospin), so the particles are indistinguishable to the eyes of the sLOCC projection. The action of projector $\hat{\Pi}^{(N)}$ on the $N$-qubit deformed state $\ket{\Psi_{D}^{(N)}}$ gives the final state 
\begin{equation}\label{Eq: N-particle sLOCC}
    \ket{\Psi^{(N)}}=\frac{1}{\sqrt{N_g}}\sum_{k}  S_k\ket{R_1\sigma_1^{(k)},\ldots,R_N\sigma_N^{(k)}},
\end{equation}
where $N_g$ is a normalization constant. The probability amplitudes $S_k=\sum_P\eta^{P}\langle \varphi_{1D}\ket{{R_{P_{1}}}\sigma_{P_{1}}^{(k)}}\dots\langle \varphi_{ND}\ket{R_{P_N}\sigma_{P_N}^{(k)}}$ are the inner products between $N$-qubit states, defined in Ref.~\cite{nosrati2020robust}, where $P=\{P_1,\dots P_N\}$ runs over all the single-qubit state permutations. The statistics factor $\eta=\pm1$ for bosons and fermions, respectively: so, $\eta^P$ is $1$ in the case of bosons and $1$ ($-1$) for even (odd) permutations in the case of fermions.

Due to the postselection, the target state $\ket{\Psi^{(N)}}$ is conditionally activated with a success (sLOCC) probability $\mathcal{P}\left(\ket{\Psi^{(N)}}\right)=\mathrm{Tr}\left(\hat{\Pi}^{(N)}\ket{\Psi_D^{(N)}}\bra{\Psi_D^{(N)}}\right)$. We highlight that the sLOCC probability, as a witness, can detect the existence of a spatial overlap between the qubits in the state when the condition of the sLOCC probability, $\mathcal{P}\neq 1$, is met, i.e., the qubits are spatially overlapping in $N$ spatial regions. Depending on how the deformation distributes the qubits, the produced target state $\ket{\Psi^{(N)}}$ can assume different entanglement structures, as we shall see in the following.

\section{Graph picture of the multipartite entangling gate}\label{sec: Result II}
It has been shown that graph theory representations offer a powerful interface for developing computer-designed quantum optics experiments, which can be applied to specific quantum information processing tasks \cite{Krenn2020,PhysRevX.11.031044,ruiz2022digital}. We now provide a graph-theoretical depiction by directly translating the three steps of the mentioned indistinguishability-based entangling gate (initialization, deformation, and activation) into a weighted bipartite graph. A (bipartite) bigraph $G_b$ consists of vertices from two disjoint sets, $V$ and $U$, where each edge $e_{ij}\in E$ connects a node $u_i\in U$ and a node $v_j\in V$ with a complex weight $r_{ij}$ \cite{west2001introduction}. In our context, each disjoint set of (quantum) nodes consists of $N$ single-qubit state vectors, given as $V=\{\ket{\varphi_{1}}, \cdots, \ket{\varphi_{N}}\}$ and $U=\{\ket{\chi_{1}}, \cdots, \ket{\chi_{N}}\}$. Those single-qubit states belonging to set $V$ ($\ket{\varphi_{i}} \in V$) consist of a wave function and an initial pseudospin, that is, $\ket{\varphi_{i}}=\ket{\psi_i\:\tau_i}$. Instead, the single-qubit states of the set $U$ ($\ket{\chi_{j}} \in U$) are related to a detection region $R_j$ where the qubit with original state $\ket{\varphi_{i}}$ can be found with any pseudospin state, that is $\ket{\chi_{j}}=\ket{R_{j}\ \sigma_j}$ ($\sigma_j=\uparrow,\downarrow$).

To establish the connectivity of the bigraph, a deformation is performed, as detailed in the previous section. This deformation, acting on a node $\ket{\varphi_{i}}$, establishes connections from the node $\ket{\varphi_{i}}$ in set $V$ to all possible nodes in set $U$ through a designated probability amplitude $r_{ij}$, as appears in the deformed single-qubit state $\ket{\varphi_{iD}} = \sum_{j=1}^{N} r_{ij}\ket{R_j\ \sigma_j}$. This scenario is equivalent to the possible detection of a single qubit in all various regions originating from the state $\ket{\varphi_{i}}$. Consequently, each edge in the bigraph $G_b$ corresponds to the situation where a qubit is detected in the state $\ket{\chi_{i}}$, which originates from the state $\ket{\varphi_{i}}$.

A balanced bigraph can be decomposed as $G_{b}=G_{b1}\cup\dots\cup G_{bN}$ into perfect matching sub-bigraph graphs $G_{bi}$ where every two node of two sets (i.e., $U$ and $V$) belongs to precisely one of the edges \cite{fukuda1994finding,tassa2012finding,uno1997algorithms}. In this picture, performing the projective measurement of Eq.~\eqref{eq: sLOCC proj}, which aims to find one qubit per region, resembles the well-known problem of finding all perfect matching sub-graphs \cite{fukuda1994finding}. Also, we can associate a weighted adjacency matrix, $\mathcal{A}_{bi}$, to each of these sub-graphs, which can be written in the form of a block matrix as
\begin{equation}\label{eq: matrix sLOCC}
    \mathcal{A}_{bi}=\begin{pmatrix}
    0 & \mathcal{R}_{\sigma_{1},\ldots,\sigma_{N}}\\
    \mathcal{R}_{\sigma_{1},\ldots,\sigma_{N}}^{T} & 0
\end{pmatrix}.
\end{equation}
Here, the entry  $\mathcal{R}_{\sigma_{1},\ldots,\sigma_{N}}$ represents the (nonzero block) weight matrix, with  $\mathcal{R}_{\sigma_{1},\ldots,\sigma_{N}}^{T}$ denoting its matrix transpose conjugate. This block weight matrix $\mathcal{R}_{\sigma_{1},\ldots,\sigma_{N}}$ indicates edges between two sets of nodes, namely $U$ and $V$, with matrix elements  $\mathcal{R}_{\sigma_{1},\ldots,\sigma_{N}}=\{\bra{\chi_i}\varphi_{jD}\rangle\}$ ($i,j=1,2,\ldots,N$). A zero matrix block element indicates the absence of an edge between nodes of each set. 
Using weight matrices corresponding to each of the perfect matches, we ultimately obtain the generated state 
\begin{equation}\label{eq: determinant of the weight matrix}
    \ket{\Psi^{(N)}}=\frac{1}{\sqrt{N_g}}\sum_{\sigma_{1},\dots\sigma_{N}=\{\uparrow,\downarrow\}}\left| \mathcal{R}_{\sigma_{1},\ldots,\sigma_{N}}\right|_{\eta}\ket{R_1\sigma_1,R_2\sigma_2,\dots,R_N\sigma_N},
\end{equation}
where $N_g$ is the normalization constant and the sum runs over all the $2^N$ pseudospin configurations.  Hence, $\left| \mathcal{R}_{\sigma_{1},\ldots,\sigma_{N}}\right|_{\eta}$ recalls the determinant-like of the weight matrix of each perfect match, taking into account the particle statistics $\eta$. Based on the Laplace approach of matrix determinant \cite{poole2003linear}, 
we can define the determinant-like as  $\left| \mathcal{R}_{\sigma_{1},\ldots,\sigma_{N}}\right|_{\eta}=\sum_{i,j=1}^N\eta^{(i+j)}r_{ij}M_{ij}$, where $\{M_{ij}\}$ are the elements of the minors of the matrix $\mathcal{R}_{\sigma_{1},\dots\sigma_{N}}$. As a matter of fact, the factor $\left| \mathcal{R}_{\sigma_{1},\ldots,\sigma_{N}}\right|_{\eta}$ represents the inner products between the deformed state and the various states of the sLOCC projection measurement $\hat{\Pi}^{(N)}$ of Eq.~\eqref{eq: sLOCC proj}, as obtained previously. The output state of Eq.~(\ref{eq: determinant of the weight matrix}) in this graph picture is thus exactly equivalent to the target state of Eq.~(\ref{Eq: N-particle sLOCC}) obtained after the action of the projection $\hat{\Pi}^{(N)}$.

Also, one can show a one-to-one correspondence between a balanced bigraph and a directed graph (digraph) \cite{fukuda1994finding,tassa2012finding,uno1997algorithms}. In this representation, the set of nodes $U=\{u_1,\dots,u_n\}$ is a binary relation between the single-qubit state and the measurement region $u_i=(\ket{\varphi_{i}},\: \ket{\chi_{i}}$). This means that the detection of a qubit with $\ket{\varphi_{i}}$ in the region $\ket{\chi_{j}}$ corresponds to a directed edge from node $u_i$ to $u_j$ with weight $r_{ij}$. In the case where $i=j$, it indicates a self-loop. Looking at it from another angle, the spatial overlap between single-qubit states in a specific region determines the connectivity of the graph. An important fact is that qubit spatially overlaps if and only if one vertex of the corresponding digraph has more than one edge.

\begin{figure}[!t]
    \centering
\includegraphics[width=0.9\linewidth]{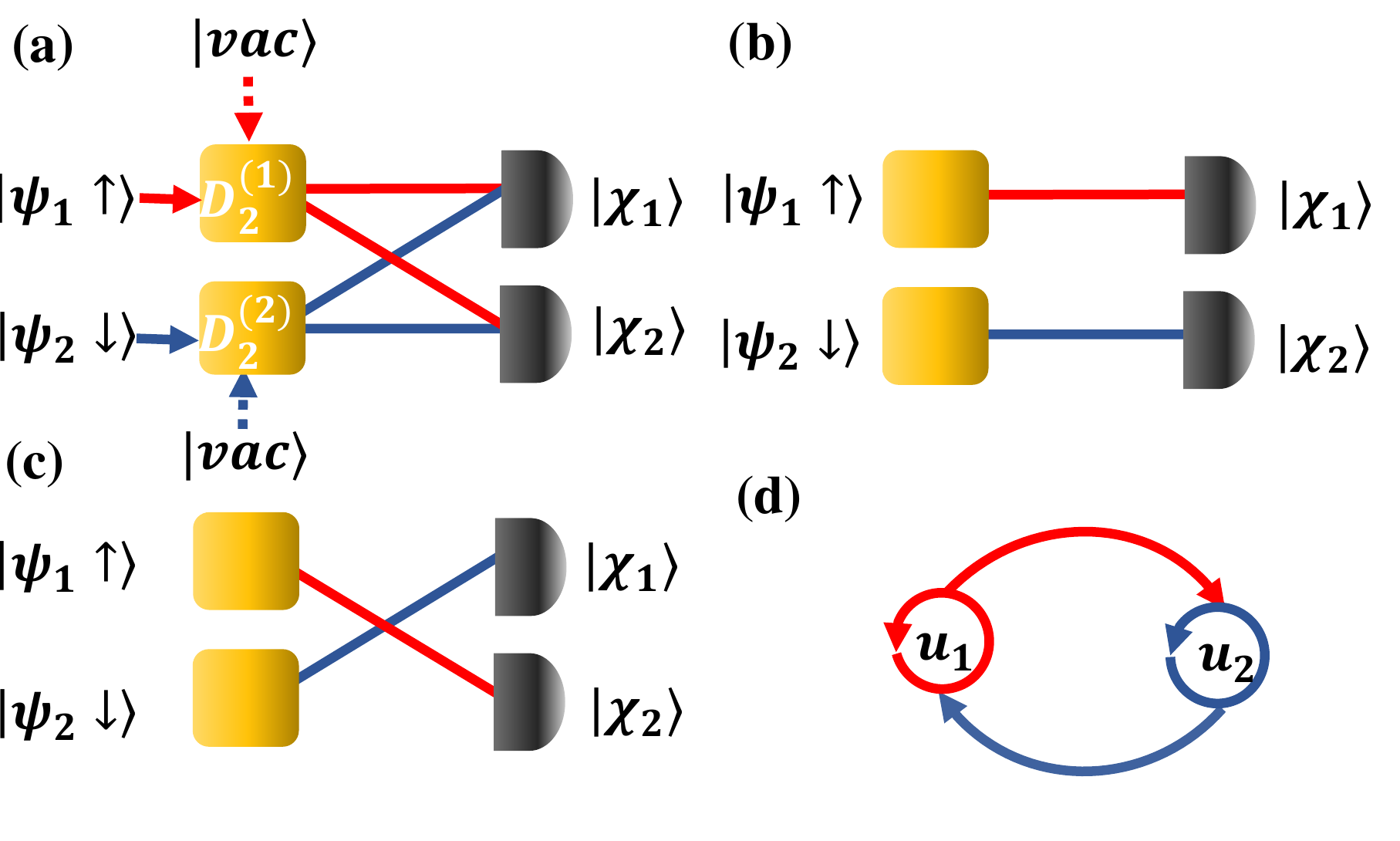}
    \caption{Graph representations of entangling protocol in the case of two particles. \textbf{(a)} bigraph with $G_b=(V\cup U,E)$, where $V=\{\ket{\psi\uparrow},\ket{\psi\downarrow}\}$ and $U=\{\ket{\chi_{1}},\ket{\chi_{2}}\}$, edge $e_{ij}\in E$ connects a node $u_i\in U$ and a node $v_j\in V$ with complex weight $r_{ij}$. Panels \textbf{(b)} and \textbf{(c)} display perfect  matching sub-graphs that correspond to states $\ket{R_1\uparrow,R_2\downarrow}$ and $\ket{R_1\downarrow,R_2\uparrow}$. \textbf{(d)} Colored digraphs (with self-loops) with nodes $U=\{u_1,u_2\}$, where each node is a binary relation between the single-qubit state and the measurement region $u_i=(\ket{\varphi_{i}},\: \ket{\chi_{i}}$). Blue and red correspond to down and up pseudospins,  respectively.}
    \label{fig: two particles}
\end{figure}

\subsection{Graph picture of bipartite entanglement generation}

As a starting point for our work, we review the bipartite entanglement generation scheme, which has been theoretically and experimentally demonstrated in Ref.~\cite{franco2018indistinguishability,sun2020experimental}, from a graph-theoretical perspective. Let us take two qubits with opposite pseudospins in a spatially separated configuration, represented by the state $\ket{\Psi_0}=\ket{\varphi_1}\otimes\ket{\varphi_2}=\ket{\psi_{1}\uparrow}\otimes\ket{\psi_{2}\downarrow}$. This initial state establishes the first set of nodes, $V=\{\ket{\psi_{1}\uparrow},\ket{\psi_{2}\downarrow}\}$. Two detection regions instead result in the second set of nodes, namely $U=\{\ket{\chi_1},\ket{\chi_2}\}$, where $\ket{\chi_j}=\ket{R_{j}\ \sigma_j}$ ($j=1,2$). As described above, a deformation is applied to the initial states of the two qubits, distributing them into the separated measurement regions $R_{1}$, $R_2$. In the remote design, the overall two-mode deformation is given by $\ket{\varphi_{1D}}=D_{2}^{(1)}\ket{\psi_{1}\uparrow}=r_{11}\ket{R_{1}\uparrow}+r_{12}\ket{R_{2}\uparrow}$ and $\ket{\varphi_{2D}}=D_{2}^{(2)}\ket{\psi_{2}\downarrow}=r_{21}\ket{R_{1}\downarrow}+r_{22}\ket{R_{2}\downarrow}$. An experimental realization of such spatial deformation has been achieved using two independent beam splitters, which distribute two photons to separate regions $R_{1}$ and $R_{2}$ without altering their polarization \cite{sun2020experimental}. After performing the sLOCC measurement with $r_{ij}=1/\sqrt{2}$, the desired Bell-state, denoted as $\ket{\Psi^{(2)}}=\left(\ket{R_{1}\uparrow,R_{2}\downarrow}+\eta\ket{R_{1}\downarrow,R_{2}\uparrow}\right)/\sqrt{2}$, which exhibits maximal entanglement for both fermionic and bosonic statistics, is generated. Furthermore, this state can also be generated using an on-site scheme with a single spatial deformation. In this approach, two qubits with opposite pseudospins pass through each mode of the deformation with different coefficients $r_{ij}$ that should satisfy the unitarity constraint of the spatial deformation.

The graph representation in Fig.~\ref{fig: two particles}~(a) illustrates the outcome of the three-step generation scheme. Here, the two sets are connected and form a bigraph, where the edge colors denote the different pseudospins of the two qubits.  In the two-qubit scenario shown in Fig.~\ref{fig: two particles}~(b), there are two perfect matching sub-graphs. The first perfect matching sub-graph involves the qubit with an up pseudospin $\ket{\psi_1\uparrow}$ detected in region one $\ket{R_1\uparrow}$ and the one with a down pseudospin $\ket{\psi_2\downarrow}$ detected in region two $\ket{R_2\downarrow}$. The second matching involves the opposite, with the qubit having a down pseudospin $\ket{\psi_2\downarrow}$ detected in region one $\ket{R_1\downarrow}$ and the one with an up pseudospin $\ket{\psi_1\uparrow}$ detected in region two $\ket{R_2\uparrow}$ (Fig.~\ref{fig: two particles}~(c)). For each of these perfect matches, we can define the weight matrix adjacency as
\begin{equation}
\mathcal{R}_{\sigma_1,\sigma_2}=
\begin{pmatrix}
    \bra{\chi_1}{\varphi_{1D}}\rangle & \bra{\chi_1}{\varphi_{2D}}\rangle\\
    \bra{\chi_2}{\varphi_{1D}}\rangle\rangle & \bra{\chi_2}{\varphi_{2D}}\rangle
\end{pmatrix},    
\end{equation}
indicating the connectivity of the perfect match in the sub-graphs. Finally, we get the output state that is in the superposition of the two mentioned perfect matching, given by
\begin{equation}
    \ket{\Psi^{(2)}}=\frac{1}{\sqrt{N_g}}\sum_{\sigma_{1},\sigma_{2}=\{\uparrow,\downarrow\}}\left| \mathcal{R}_{\sigma_1,\sigma_2}\right|_{\eta}\ket{R_1\sigma_1,R_2\sigma_2},
\end{equation}
where the determinant-like operation $\left| \mathcal{R}_{\sigma_1,\sigma_2}\right|_{\eta}=\bra{\chi_1}{\varphi_{1D}}\rangle\bra{\chi_2}{\varphi_{2D}}\rangle+\eta\bra{\chi_1}{\varphi_{2D}}\rangle\bra{\chi_2}{\varphi_{1D}}\rangle\rangle$ is precisely the result that has been identified in Ref.~\cite{Lo_Franco2016Quantum} as the inner product between states of two identical qubit (two-qubit probability amplitude). One can easily see that the expansion of the above expression is equal to $\ket{\Psi^{(2)}}=\left({r_{11}r_{22}\ket{R_{1}\uparrow,R_{2}\downarrow}+\eta r_{21}r_{12}\ket{R_{1}\downarrow,R_{2}\uparrow}}\right)/\sqrt{N_g}$ where $N_g=\vert r_{11}r_{22}\vert^{2}+\vert r_{21}r_{12}\vert^{2}$. It can be observed that the Bell state can be generated with $r_{ij}=1/\sqrt{2}$ choices and the sLOCC probability $\mathcal{P}\left(\ket{\Psi^{(2)}}\right)=1/2$ sLOCC probability. Also, we illustrate the colored digraph of detecting two qubits in two regions in Fig.~\ref{fig: two particles}~(d), where red and blue edges represent the up and down pseudospins, respectively. In appendix~\ref{App: Bell State}, we demonstrate the generation of another type of Bell state, namely $\ket{\Phi_{+}^{(2)}}=\frac{1}{\sqrt{2}}(\ket{R_{1}\uparrow,R_{2}\uparrow}+\ket{R_{1}\downarrow,R_{2}\downarrow})$. This is accomplished using a general form of deformation that not only alters the spatial wave functions but also modifies the internal degrees of freedom.

\section{Multipartite entanglement generation schemes}
In the following, we provide indistinguishability-based generation schemes for genuine multipartite entangled states using a graph-theoretical representation approach. The multipartite states of interest are W, Dicke, GHZ, and cluster states.

\subsection{$W$ state} 
We consider a $W$ state to be the first type of genuine multipartite entangled state. The $W$ state is an equal combination of all potential pure states, where only one of the qubits has pseudospin $\uparrow$ while the remaining qubits have pseudospins $\downarrow$. We assume that each identical qubit will ultimately occupy distinct spatial regions. Therefore, an $N$-qubit $W$ state of identical qubits has the form
\begin{equation}\label{eq: $W$ state N}
    \begin{split}
    \ket{W^{(N)}}&=\frac{1}{\sqrt{N}}\Big(\ket{R_1\uparrow,R_2\downarrow,\dots,R_N\downarrow}+\ket{R_1\downarrow,R_2\uparrow,\dots,R_N\downarrow}
    \\
    &+\cdots+\ket{R_1\downarrow,R_2\downarrow,\dots,R_N\uparrow}\Big).
    \end{split}
\end{equation}
To generate an $N$-qubit $W$ state, we start with the pure product state of $N$ qubits as $\ket{\Psi_0}=\ket{\psi_{1}\uparrow}\otimes\ket{\psi_{2}\downarrow}\otimes\dots\otimes\ket{\psi_{N}\downarrow}$, where $N-1$ qubits have pseudospin down ($\downarrow$) and one qubit has pseudospin up ($\uparrow$). The next step involves designing the deformation, either in remote or on-site schemes, as illustrated in Fig.~\ref{fig: deformation illustration}.

\begin{figure}[!t]
    \centering
\includegraphics[width=0.9\linewidth]{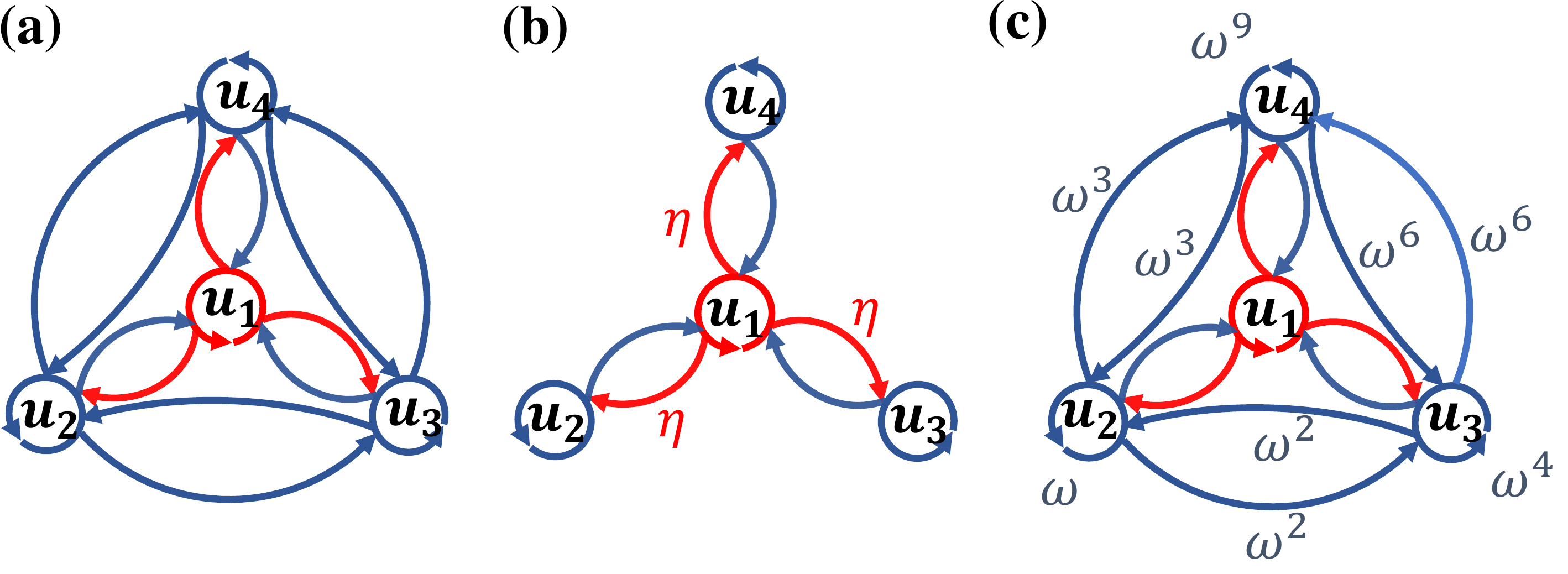}
    \caption{\textbf{Digraph representations of four-particle $W$ state generation.} The digraph illustrates four quantum nodes $u_j$ ($j=1,\dots,4$), representing a binary state and measurement region. Red (blue) edges indicate the detection of a qubit with an up (down) pseudospin. The presence of $\eta$ on red edges denotes a $\pi$ phase shift between connected nodes for fermions. \textbf{(a)} The complete digraph illustrates mutually oriented edges between every pair of nodes, each with equal weighting. \textbf{(b)} The star graph displays mutually oriented edges only between the center node and the other nodes. \textbf{(c)} The QFT digraph exhibits mutually oriented edges between each pair with a specific weighting as $\omega_N^n$ ($\omega_N=e^{2i\pi/N}$).}
    \label{fig: W graph}
\end{figure}

As a first remote design, we consider a deformation operation that equally distributes each qubit among $N$ regions, expressed as $\hat{D}_{N}^{(i)}\ket{\psi_{i}\sigma_i}=\ket{\varphi_{iD}}=\frac{1}{\sqrt{N}}\sum_{j=1}^{M}\ket{R_j\:\sigma_{i}}$. Subsequently, with the action of the projection $\hat{\Pi}^{(N)}$ (Eq.~\eqref{eq: sLOCC proj}), the $W$ state (Eq.~\eqref{eq: $W$ state N}) is generated. Such a generation scheme can be interpreted using a complete digraph (digraph) with equal weight, denoted as $G_{C}$ (where $C$ represents the complete digraph). In a complete digraph, each pair of graph vertices is connected by a pair of edges with non-zero weights (one in each direction), as illustrated in Fig. \ref{fig: W graph}~(a) for four qubits. The perfect matching can be found using Eq.~(\ref{eq: determinant of the weight matrix}), and the $W$ state (Eq.~\eqref{eq: $W$ state N}) is a superposition of these perfect matchings (see {Appendix~\ref{appx W state}} for more details on three- and four-qubit states). The complete digraph with equal weight designed to generate the bosonic $W$ state (Eq.~\eqref{eq: $W$ state N}) has the success (sLOCC) probability given by
\begin{equation}\label{eq: prob $W$ state Complete Graph}
    \mathcal{P}_{\mathrm{C}}\left({\ket{W^{(N)}}}\right)=\frac{(N-1)!}{N^{(N-1)}}.
\end{equation}
However, generating the fermionic $W$ state with the complete digraph $G_{C}$ is impossible due to the zero sLOCC probability, $\mathcal{P}_{\mathrm{C}}\left({\ket{W^{(N)}}}\right)=0$. This zero probability can be attributed to the Pauli exclusion principle \cite{pauli2012general}, which prohibits two fermions with the same pseudospin from occupying the same spatial wave function. Therefore, an alternative deformation scheme is needed to generate the fermionic $W$ state.

An alternative remote approach to generating the $W$ state involves employing the generation scheme based on the star digraph $G_S$. In this configuration, as illustrated in Fig.~\ref{fig: W graph}~(b) for four nodes, a qubit with the opposite pseudospin (i.e., up) resides in the center and spatially overlaps with all other qubits, while the remaining qubits with a down pseudospin only spatially overlap with the central one. To construct such a generation scheme, spatial deformation operations are defined as $\hat{D}_{N}^{(1)}\ket{\psi_{1}\uparrow}=\ket{\varphi_{1D}}=\frac{1}{\sqrt{N}}\left(\ket{R_1\uparrow}+\eta\sum_{i=2}^{M}\ket{R_i\uparrow}\right)$ and $\hat{D}_{N}^{(i)}\ket{\psi_{i}\downarrow}=\ket{\varphi_{iD}}=\frac{1}{\sqrt{2}}(\ket{R_1\downarrow}+\ket{R_i\downarrow})$, for $i=2,\dots, N$, where $\eta=\pm1$. It is worth mentioning that the generation of the $W$-state depends not only on the form of deformation but also on particle statistics. When designing a generation scheme, one must consider the particle exchange phase $e^{i\phi}$ ($\phi=0$ for bosons and $\phi=\pi$ for fermions) that emerges when two single-qubit states are swapped \cite{peres1997quantum}. Finally, after applying a projection measurement, we can generate the $W$ state for both bosonic and fermionic statistics. Interestingly, the sLOCC probability for fermions is given by  
\begin{equation}\label{eq: prob of N star}
\mathcal{P}_{star}\left(\ket{W^{(N)}}\right)=\frac{1}{N},
\end{equation}
which is significantly enhanced compared to the sLOCC probability in the complete digraph case for bosons. 

Furthermore, one can employ on-site deformation, where each identical qubit is sent through one of the input modes, as illustrated in Fig.~\ref{fig: deformation illustration}~(a). In this scenario, the initial state undergoes a spatial quantum Fourier transform (QFT) \cite{nielsen2010quantum}. For each initial single-qubit state $\ket{\psi_j\:\sigma}$, we can express the QFT using the following map:
\begin{equation}\label{eq: QFT}
    \ket{\psi_j\:\sigma}\mapsto\frac{1}{\sqrt{N}}\sum_{k=1}^{N}\omega_N^{(j-1)(k-1)}\ket{R_k\:\sigma},
\end{equation}
where $\omega_N=e^{2i\pi/N}$ is the relative phase. Here, the QFT deformation does not alter the pseudospin $\sigma$. This deformation represents a special form of a complete graph with specific complex weight—a relative phase $\omega_N$ introduced by the QFT—as illustrated in Fig.~\ref{fig: W graph}~(c) for four qubits. Finally, the $W$ state can be generated for both fermions and bosons (only in the case of an odd number of qubits for bosons; see appendix~\ref{appx W state}) using such QFT-based on-site design. The sLOCC probability in the generation of the $W$ state with fermions using the QFT-based digraph representation is given by 
\begin{equation}
\mathcal{P}_{QFT}\left(\ket{W^{(N)}}\right) = \frac{1}{N},
\end{equation}
which is equivalent to the sLOCC probability of $W$ state generation using the star digraph representation in the remote design. {It is worth mentioning that the star digraph representation for W-state generation, as illustrated in Fig. 3, has been demonstrated in Ref \cite{chin2021graph} via an optical linear on-site design. On-site design is based on the linear transformation of creation (annihilation) operators in the second quantization formalism.}

\begin{figure}[t!]
    \centering
\includegraphics[width=1.0\linewidth]{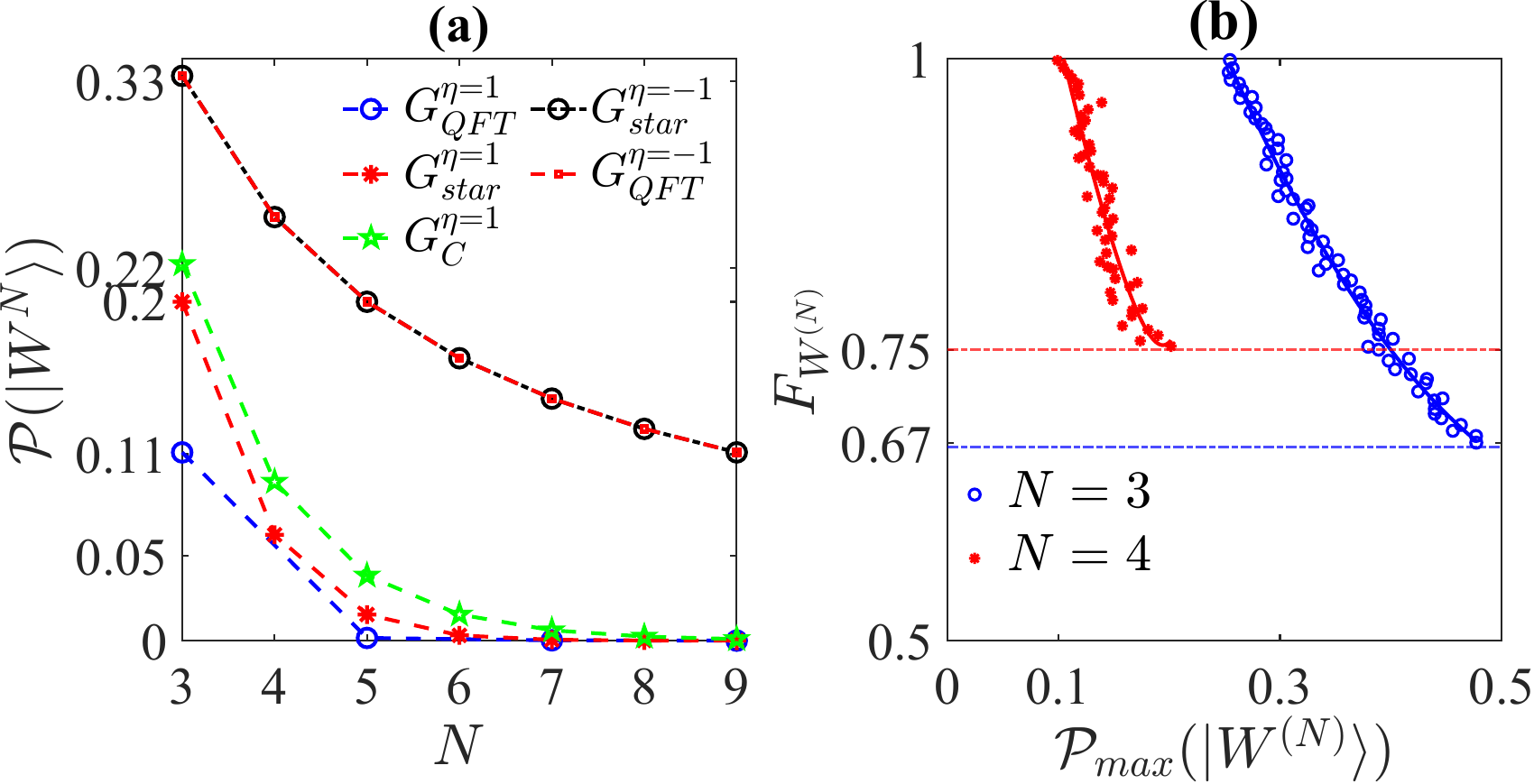}
    \caption{\textbf{Success probability and fidelity of generating $W$ states with different schemes and particle statistics.} \textbf{(a)} Success probability $\mathcal{P}(\ket{W^{(N)}})$ as a function of the number of particles $N$ for different $W$ state generation schemes, including the complete digraph for bosons (green star), the star digraph for bosons (red asterisk), the QFT digraph for bosons (blue circle), the star digraph for fermions (black circle), and the QFT digraph for fermions (red square). \textbf{(b)} Fidelity as a function of the maximum probability of success $\mathcal{P}_{\text{max}}(\ket{W^{(N)}})$ for three bosons (blue circles) and four bosons (red stars) in the form of genuine $W$ states in the complete digraph. The dashed lines correspond to the bound that guarantees genuine multipartite $W$ entanglement.}
    \label{fig: trade-off $W$ state}
\end{figure}

In order to recap, we compare different possible schemes in terms of sLOCC probability to assess the generation scheme efficiency. We plot the sLOCC probability for generating an $N$-qubit $W$ state as a function of the number of qubits $N$ in Fig.~\ref{fig: trade-off $W$ state}~(a) using complete, star, and QFT digraphs with both particle statistics. As depicted in Fig.~\ref{fig: trade-off $W$ state}~(a), fermions exhibit better performance in both on-site and remote schemes. This advantage can be attributed to the Pauli exclusion principle, which prohibits fermions from coexisting in the same region with the same pseudospin. Moreover, different configurations allow for many different structures of multipartite (refer to appendix~\ref{appendix: Dicke state} for more information on the general form of generating three- and four-qubit states). {Furthermore, the weight matrix adjacency in a star digraph can be adjusted to maximize the success probability, analogous to what was reported in Ref. \cite{PhysRevA.104.023701} as an efficient optical setup to generate a W state. It is worth noting that the graph-theoretic approach enables the discovery of alternative generation schemes for W states, as can be seen in Refs. \cite{Quantumexperimentsandgraphs22017Xuemei,chin2021graph} with bosonic statistics.} 

We have already demonstrated that the ideal scenarios of complete, star, and QFT digraphs can lead to the generation of the $W$ state. However, it is essential to explore the imperfect spatial overlap configurations where probability amplitudes $r_{ij}$ (weight of the edges in digraphs) are not exactly our desired ones. When we have an imperfect situation, the crucial question is whether the generated state is genuinely multipartite-entangled. A typical figure of merit is fidelity, which measures how close the generated state is to an ideal pure entangled state \cite{PhysRevA.76.030305, Campbell_2009,Toth:07, PhysRevA.74.020301}. Formally, fidelity is a measure of the distance from a general generated state $\rho$ to the desired pure state $\ket{\psi}$, given by $F_\psi=\bra{\psi}\rho\ket{\psi}$. Therefore, if the fidelity between the generated state and the multipartite $W$ state violates the inequality $F_W\leq\frac{N-1}{N}$, the generated state is genuinely an $N$-partite (here $N$-qubits) $W$ state \cite{Haffner2005,PhysRevLett.92.087902}. 

Besides the imperfections of the generation scheme, it is crucial to assess the maximum probability of success since our scheme is probabilistic. Such a consideration leads to an interesting trade-off between fidelity and the probability of success, in which one may sacrifice perfect entanglement generation for a higher probability of success. Here, the objective is to maximize the success probability, denoted as $\mathcal{P}(\ket{W^{(N)}})$, by searching for an optimal digraph, i.e., a generation scheme, while ensuring that the fidelity does not fall below a certain threshold for the multipartite $W$ state, specifically $F_{W}\leq \frac{N-1}{N}$. To solve the maximization problem, we first select the complete digraph configuration with bosonic statistics. Then, we randomly sample a large number of digraphs with positive weights and calculate both the fidelity and sLOCC probability. Subsequently, we divide the fidelity values exceeding the threshold $F_W\leq\frac{N-1}{N}$ into small intervals and determine the maximum sLOCC probability for each of these intervals. As a result, we plot the fidelity as a function of its maximum probability of success for three and four bosons in Fig.~\ref{fig: trade-off $W$ state}~(b). Additionally, for a digraph with all possible edges, the highest sLOCC probabilities that guarantee the generation of genuine three- and four-boson $W$ states, respectively, are $\mathcal{P}_3=0.47$ and $\mathcal{P}_4=0.20$.

\subsection{Symmetric Dicke state} 

The Dicke state is another type of multipartite entangled state, which was first investigated concerning light emission from clouds of atoms \cite{dicke1954coherence}. We assume that each qubit will reside in different detection regions $R_j$. Thus, the $N$-qubit symmetric Dicke state with $N/2$ down pseudospins can be defined as follows:
\begin{equation}\label{eq: Dicke sLOCC}
     \begin{split}
         \ket{D^{(N)}}=\binom{N}{\frac{N}{2}}^{-\frac{1}{2}}\sum_{P_j}^{N}|&R_{1}\uparrow_{P_1},\dots,R_{N/2}\uparrow_{P_{N/2}},
     \\
     &R_{N/2+1}\downarrow_{P_N/2+1},\dots, R_{N}\downarrow_{P_N}\rangle,
     \end{split}
\end{equation}
where the summation goes over all distinct permutations of pseudospins while keeping the regions $R_i$ fixed.

\begin{figure}[!t]
    \centering
\includegraphics[width=1.0\linewidth]{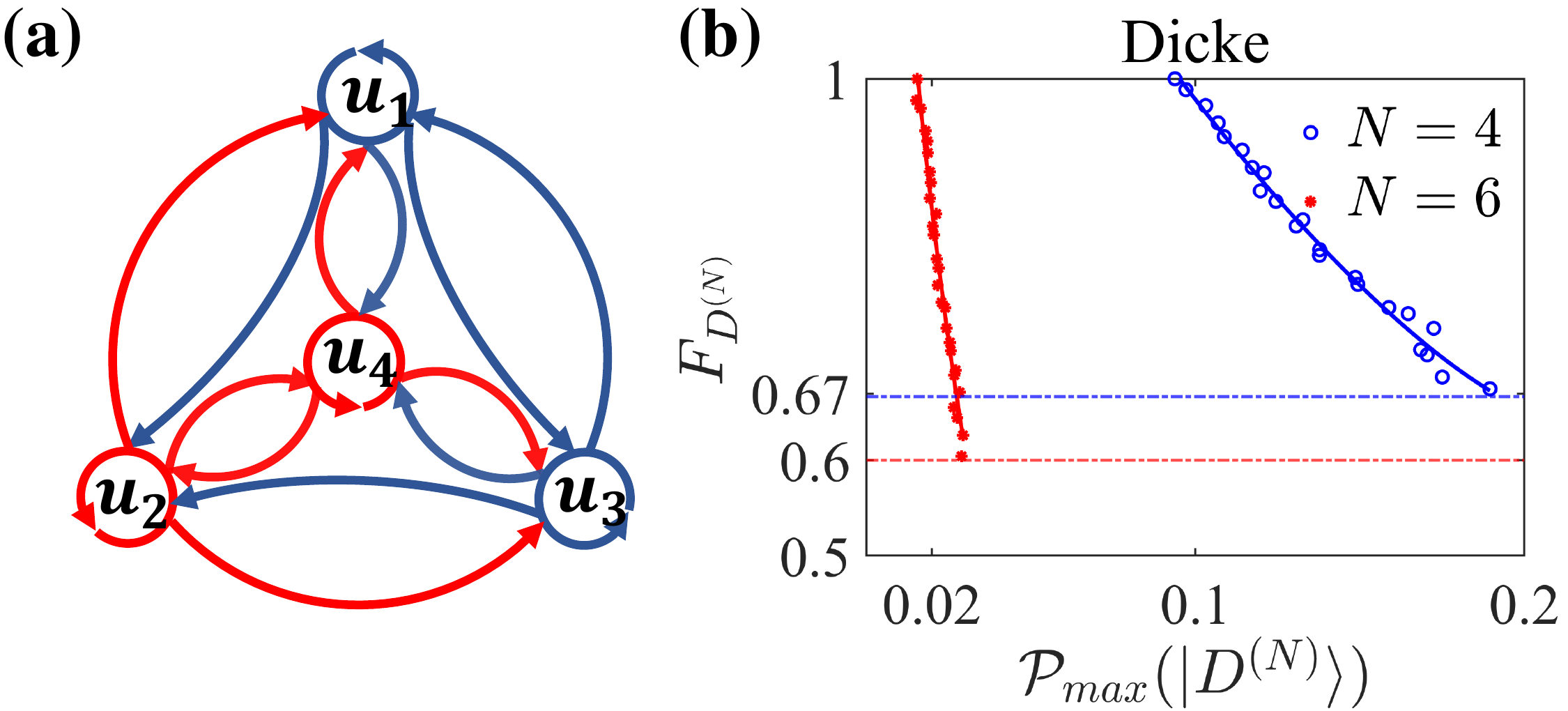}
    \caption{\textbf{ {Symmetric} Dicke state generation.} \textbf{(a)} Colored complete digraph $U=\{u_1,\dots u_4\}$ for a four-qubit {symmetric} Dicke state. \textbf{(b)} Fidelity as a function of maximum probability of success {$\mathcal{P}_{max}(\ket{D^{(N)}})$} for four (blue circles) and six (red stars) bosons in the form of a genuine {symmetric} Dicke state with two pseudospins down in the complete digraph. The dashed lines correspond to the bound that guarantees genuine multipartite symmetric Dicke entangled states.}
    \label{fig: Dicke state}
\end{figure}

We start the generation process with the pure initial state of an $N$-qubit system as $\ket{\Psi_0}=\ket{\psi_{1}\uparrow}\otimes\dots\otimes\ket{\psi_{N/2}\downarrow}\dots\ket{\psi_{N/2+1}\uparrow}\otimes\dots\otimes\ket{\psi_{N}\downarrow}$, which comprises half of the qubits with up pseudospin and the other half with down pseudospin. As the first generation scheme, we employ a remote design and apply a spatial deformation to each qubit as $\hat{D}_{N}^{(i)}\ket{\psi_{i}\sigma_{i}}=\ket{\varphi_{iD}}=\frac{1}{\sqrt{N}}\sum_{j=1}^{M}\ket{R_j\:\sigma_{i}}$, which distributes each qubit equally among all detection regions. After applying the projection measurement, the bosonic symmetric-Dicke state is generated. Such a generation process is based on the complete digraph with equal weights, as illustrated in Fig.~\ref{fig: Dicke state}~(a) for four qubits. The associated sLOCC probability in generating the $N$ bosonic symmetric-Dicke state is given by: 
\begin{equation}\label{eq: sLOCC probability Dicke}
\mathcal{P}_{C}\left(\ket{D^{(N)}}\right)=\frac{(N-1)!}{N^{N-1}}.
\end{equation}
{Alternatively, one can find different digraphs that still lead to the generation of symmetric Dicke states. For example, by removing connections in complete graphs, one can check whether a symmetric Dicke state is generated. Notably, alternative remote designs for generating four qubits in bosonic symmetric Dicke states have been introduced in Ref. \cite{chin2021graph}.}
It is worth noting that the on-site approach with the QFT scheme is incapable of generating possible symmetric Dicke states, as reported in Ref. \cite{Scalable2018Kasture} [for details, refer also to appendix~\ref{appendix: Dicke state}].

In non-ideal situations, we can employ fidelity-based criteria to test the entanglement between $N$ identical qubits. An entangled symmetric Dicke state is a genuine $N$-partite entangled state if and only if the inequality $F_{D}\leq\frac{N}{2(N-1)}$ is violated \cite{Toth:07}. This also allows us to explore the trade-off between fidelity and the probability of success in the genuine multipartite Dicke state. We sample a large number of random digraphs and maximize the sLOCC probability for each value of $F_{D}$ over the threshold. As a result, the fidelity is plotted as a function of the maximum probability of success for four and six bosons in Fig.~\ref{fig: Dicke state}~(b). For a complete digraph with non-equal weighted edges, the maximum probability of success for genuine four-boson and six-boson Dicke states is $\mathcal{P}(\ket{D^{(4)}})=0.19$ and $\mathcal{P}(\ket{D^{(6)}})=0.03$, respectively.

\begin{figure}[!t]
    \centering
\includegraphics[width=0.9\linewidth]{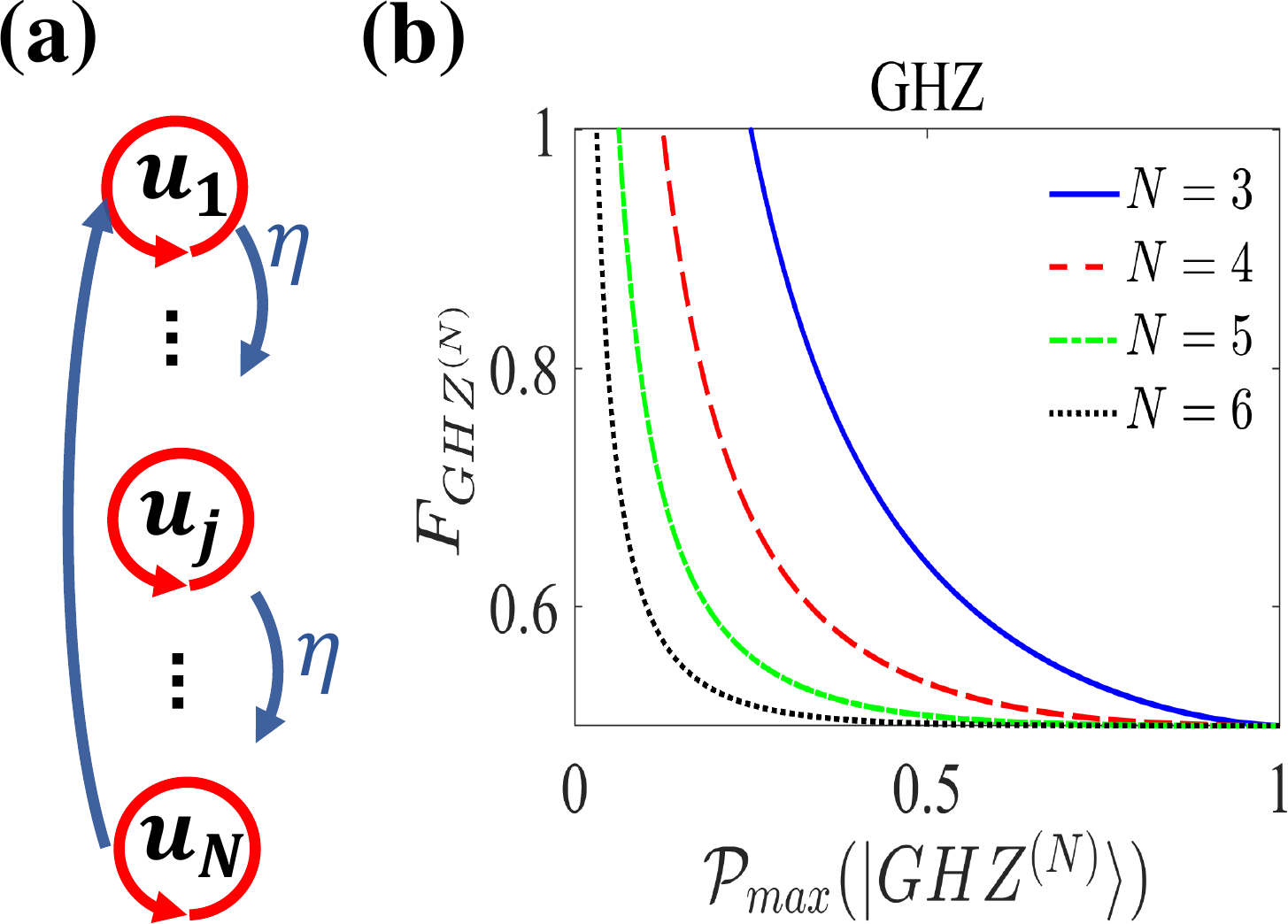}
    \caption{\textbf{(a) Digraph representations of $N$-qubit GHZ state.} Colored digraph (with self-loops) with nodes $U=\{u_1,u_2\dots u_N\}$. Also, the self-loops and oriented edges have different colors. The presence of $\eta$ shows a $\pi$ phase shift between connected nodes for fermions. Indeed, this phase shift is not necessary when working with bosons. \textbf{(b) Trade-off between fidelity and sLOCC probability.} Fidelity as a function of maximum probability of success $\mathcal{P}_{max}(\ket{GHZ^{(N)}})$ for three (blue), four bosons (red), five (green) and six (black) bosons in the form of genuine GHZ state}
    \label{fig: GHZ state}
\end{figure}

\subsection{GHZ state}

GHZ states play a central role in quantum computing \cite{zhou2022double, PhysRevLett.115.020502}, and metrology \cite{Tóth_2014}. A GHZ state is a superposition of all qubits being in either up or down pseudospin states. Assuming each qubit resides in a different spatial region, the $N$-qubit GHZ state is expressed as 
\begin{equation}\label{GHZ slocc}
    \begin{split}
    \ket{\mathrm{GHZ}^{(N)}}=\frac{1}{\sqrt{2}}(\ket{R_{1}\uparrow,\dots,R_{N}\uparrow}+\ket{R_{1}\downarrow,\dots,R_{N}\downarrow}).
\end{split}
\end{equation}

To generate a GHZ state, we begin with $N$ identical qubits in a pure product, represented as $\ket{\Psi^{(N)}}=\ket{\varphi_{1}}\otimes \ket{\varphi_{2}} \otimes \cdots \otimes \ket{\varphi_{N}}$, where $\ket{\varphi_{i}}=\ket{\psi_{i}\uparrow}$. Spatial deformation alone is insufficient for generating GHZ states, so we employ a more general form of deformation that modifies both the spatial wave function and the pseudospins of the initial qubits. In this deformation, each qubit is distributed among two regions, and its pseudospin is simultaneously altered if the qubit is detected in the nearest neighboring region. Specifically, the action of each deformation on the initial qubits is given by $\hat{D}_N^{(i)}\ket{\psi_i\uparrow}=\ket{\varphi_{iD}} = \frac{1}{\sqrt{2}}(\ket{R_{i}\uparrow} + \eta\ket{R_{i+1}\downarrow})$ for the first to $(N-1)$ qubits. Additionally, the $N$-th qubit undergoes deformation as $\hat{D}_{N}^{(N)}\ket{\psi_N\uparrow}=\ket{\varphi_{ND}} = \frac{1}{\sqrt{2}}(\ket{R_{N}\uparrow} + \ket{R_{1}\downarrow})$. After selecting one qubit per region (which is the action of $\hat{\Pi}^{(N)}$ (Eq.~\eqref{eq: sLOCC proj})), the GHZ state is generated.

In the digraph representation of the GHZ state generation scheme, as illustrated in Fig.~\ref{fig: GHZ state}~(a), we have $N$ nodes with the same self-loop edges in red and outgoing directed edges in blue that connect the $i$-th node to the $(i+1)$-th node. The differences in color among the self-loops, along with their oriented edges, indicate that the same qubit with different pseudospins will be detected in the nearest neighboring regions. {The digraph representation of the generation of the bosonic GHZ state, as shown in Fig. 6, was originally introduced in Ref. \cite{chin2021graph}. This graph representation corresponds to an experimental implementation of a three-photon GHZ state using a free-space linear optics system, as described in Ref. \cite{Lee:22}.} For the fermionic GHZ state, a $\pi$-phase is necessary for the generation of the GHZ state. Finally, the GHZ state is generated with the sLOCC probability of
\begin{equation}
    \mathcal{P}{(\ket{\mathrm{GHZ}^{(N)}})}=\frac{1}{2^{N-1}},
\end{equation}
for both bosonic and fermionic statistics.

Additionally, we discuss the trade-off between the success probability $\mathcal{P}{(\ket{GHZ^{(N)}})}$  in the generation of the GHZ state and the fidelity-based entanglement criteria $F_{GHZ} \geq \frac{1}{2}$ \cite{PhysRevA.76.030305,PhysRevLett.94.060501}. The fidelity criteria ensure that the generated state is entangled in the form of GHZ. We plot the fidelity $F_{GHZ}$ as a function of the maximum probability of success for three to six qubits (bosons and fermions) in Fig.~\ref{fig: GHZ state}~(b). The plot clearly demonstrates that one can generate a weak GHZ state with a high probability of success.

\subsection{Cluster state} 

Cluster states are a type of multipartite entangled state that serves as a fundamental building block for measurement-based quantum computing and one-way quantum computing, in which measurements on qubits within the cluster state are used to perform universal quantum computing \cite{Walther2005, Raussendorf2001One}. These classes of multipartite entangled states exhibit robustness against noise and decoherence \cite{PhysRevA.71.032350, Walther2005}. Historically, the generation of a cluster state has been based on next-neighbor interaction between qubits on a chain \cite{Walther2005, Raussendorf2001One}. In this work, we employ the indistinguishability-based entangling gate to generate the cluster state. By assuming that each qubit resides in a distinct spatial region, an $N$-qubit linear cluster can be expressed in the form
\begin{equation}
    \ket{C^{(N)}}=\frac{1}{2^{\frac{N}{2}}}\bigotimes_{i=1}^{N}\left(\ket{R_i\downarrow}\sigma_z^{R_{i+1}}+\ket{R_i\uparrow}\right).
\end{equation}
Here, the spatial Pauli operator is defined as $\sigma_{z}^{R_{i}}=\ket{R_{i}}\bra{R_{i}}\otimes\sigma_{z}$, where $\sigma_{z}=\ket{\downarrow}\bra{\downarrow}-\ket{\uparrow}\bra{\uparrow}$, and $\sigma_{z}^{R_{N+1}}=1$. The above state satisfies the cluster state condition, i.e.,  $\sigma_{x}^{R_{i}}\sigma_{z}^{R_{i+1}}\ket{C^{(N)}}=\pm\ket{C^{(N)}}$ with $\sigma_{x}^{R_{i}}=\ket{R_{i}}\bra{R_{i}}\otimes\left(\ket{\uparrow}\bra{\downarrow}+\ket{\downarrow}\bra{\uparrow}\right)$
under a unitary transformation on one or more of the qubits \cite{Walther2005, Raussendorf2001One}. Furthermore, one can consider an even number of qubits and arrive at the following state:
\begin{equation}\label{eq: sLOCC cluster}
    \begin{split}
        \ket{C^{(N)}}=
        &\frac{1}{2}\Big(\ket{R_{1}\uparrow,\dots,R_{N}\uparrow}-\ket{R_{1}\downarrow,\dots,R_{N}\downarrow}
        \\
        +&\ket{R_{1}\uparrow,\dots,R_{\frac{N}{2}}\uparrow,R_{{\frac{N}{2}}+1}\downarrow,\dots+R_{N}\downarrow}
        \\+&\ket{R_{1}\downarrow,\dots,R_{\frac{N}{2}}\downarrow,R_{{\frac{N}{2}}+1}\uparrow,\dots,R_{N}\uparrow}
        \Big).
    \end{split}
\end{equation}

To generate the cluster state in the form described above, we consider an $N$-qubit pure product state of $\ket{\Psi^{(N)}} = \ket{\varphi_{1}} \otimes \ket{\varphi_{2}} \otimes \cdots \otimes \ket{\varphi_{N}}$, where $\ket{\varphi_{i}} = \ket{\psi_{i}\uparrow}$. The generation process involves placing the $i$-th qubit in a superposition of two spatial regions using the deformation operation, given as $\hat{D}_N^{(i)}\ket{\psi_i\uparrow}=\ket{\varphi_{iD}}=\frac{1}{\sqrt{2}}\left(\ket{R_{i}\uparrow}+\ket{R_{i+1}\downarrow}\right)$, with the exception of the middle $\frac{N}{2}$ and last $N$ qubits. Subsequently, the $\frac{N}{2}$~th qubit undergoes the deformation operation as $\hat{D}_N^{(N/2)}\ket{\psi_i\uparrow}=\ket{\varphi_{\frac{N}{2}D}}=(\ket{R_{\frac{N}{2}}\uparrow}+\ket{R_{\frac{N}{2}+1}\downarrow}\mp\eta^{\frac{N}{2}}\ket{R_{N}\uparrow})/\sqrt{3}$ where the positive (negative) sign is for bosons (fermions). Finally, the deformation on the $N$th qubit is given by $\hat{D}_N^{(N)}\ket{\psi_i\uparrow}=\ket{\varphi_{ND}}=\left(\mp\ket{R_{1}\downarrow}\pm\eta^{\frac{N}{2}}\ket{R_{\frac{N}{2}+1}\downarrow}+\ket{R_{N}\uparrow}\right)/\sqrt{3}$.

\begin{figure}[!t]
    \centering
\includegraphics[width=1\linewidth]{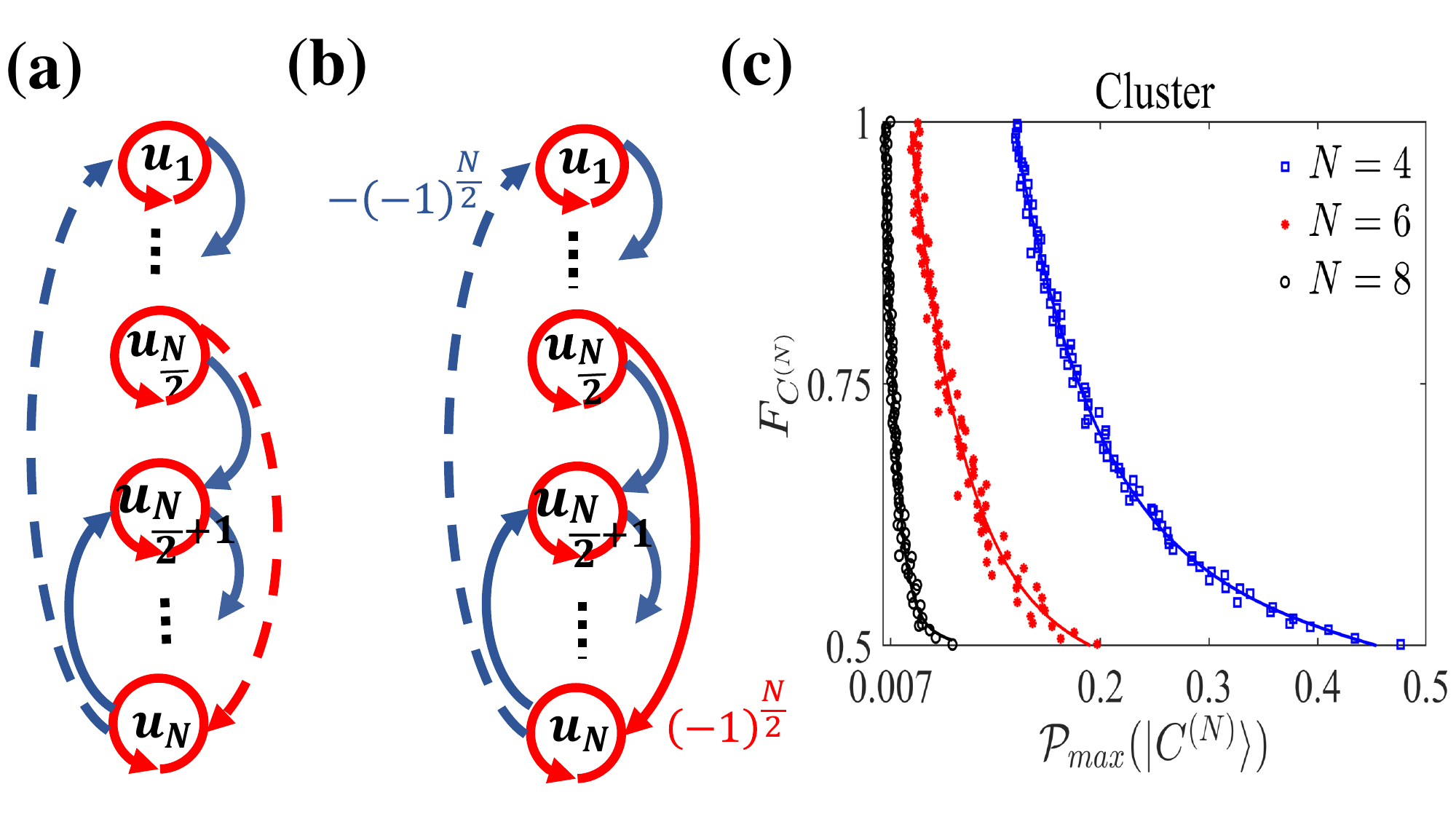}
    \caption{\textbf{Graph representations of $N$-qubit cluster states for (a) bosons and (b) fermions.} The colored digraph (with self-loops) has nodes $U=\{u_1, u_2, \ldots, u_N\}$. (a) In the bosonic design, the dashed-oriented edge represents a $\pi$ phase shift between connected nodes. (b) For fermions, the dashed-oriented edge represents a $\pm(-1)^{N/2}$ phase shift between connected nodes, depending on the number of particles. \textbf{(c) Trade-off between fidelity and sLOCC probability.} Fidelity as a function of maximum probability of success {$\mathcal{P}_{max}(\ket{C^{(N)}})$} for four (blue square), six bosons (red stars), and eight (black circle) qubits in the form of a genuine cluster state. The solid lines correspond to the fitted data.}
    \label{fig: cluster state}
\end{figure}

The digraph representation of cluster state generation, as illustrated in Fig.~\ref{fig: cluster state}~(a), consists of $N$ nodes with self-loop edges in red and outgoing directed edges in blue connecting the $i$-th node to the $(i+1)$-th node. Additionally, the node at the middle, $R_{\frac{N}{2}}$, has an additional outgoing red directed edge to the final node at $R_{N}$ with weights of $-1/\sqrt{3}$ for bosons and $e^{iN\pi/2}/\sqrt{3}$ for fermions. The node at the last phase has two outgoing directed edges in blue to $R_{\frac{N}{2}}$ and the first node $R_1$ with weights of $-1/\sqrt{3}$ for bosons and $-e^{iN\pi/2}/\sqrt{3}$ for fermions. Subsequently, the ideal cluster state in Eq.~\eqref{eq: sLOCC cluster} is generated for bosonic and fermionic statistics with the sLOCC probability given by
\begin{equation}
\mathcal{P}{(\ket{C^{(N)}})}=\frac{1}{9}\frac{1}{2^{N-4}},
\end{equation}
for an even number of qubits larger than four. 

Similar to previous sections, one might be interested in the trade-off between sLOCC probability and the amount of entanglement that can be generated using this design. Based on fidelity-based criteria, any generated state $\rho$ that violates the inequality $F_{C^{(N)}}\le \frac{1}{2}$ is considered a genuine $N$-qubit cluster state \cite{PhysRevA.76.030305,PhysRevLett.94.060501}. Additionally, to illustrate the trade-off between fidelity and the sLOCC probability, we maximize the probability of success $\mathcal{P}(\ket{C^{(N)}})$ subject to a fidelity constraint $F_{C^{(N)}}\ge \frac{1}{2}$ by sampling random positive weights over the graph representation of the cluster state {for bosonic and fermionic, respectively}, as illustrated in Fig.~\ref{fig: cluster state}~(a) and {Fig.~\ref{fig: cluster state}~(b)}.
Moreover, we plot the fidelity as a function of the maximum sLOCC probability for four to {eight} qubits (both bosons and fermions) in {Fig.~\ref{fig: cluster state}~(c)}. We obtain the maximum sLOCC probabilities $\mathcal{P}(\ket{C^{(4)}})=0.48$, $\mathcal{P}(\ket{C^{(6)}})=0.20$, and $\mathcal{P}(\ket{C^{(8)}})=0.07$ for four, six, and eight qubits (for both bosonic and fermionic statistics), respectively.

{\section{Discussion on Experimental Feasibility}
In the realm of free-space quantum optics, a few experiments have been conducted to exploit the indistinguishability of identical particles, specifically spatially overlapping photons, for generating entangled states. These experiments have successfully generated Bell states \cite{sun2020experimental,barros2020entangling}, as well as three-photon W and GHZ states \cite{Lee:22}. The internal degree of freedom, or pseudospin, of the photon is represented by the polarization state, while the spatial wave functions correspond to the regions where the photons are detected. In the following, we provide an overview of the experimental platforms needed for such experiments, as well as the potential experimental implementation of the three-step generation proposed in our work.

In the first step (initialization), one can prepare an initially separable state through various methods. One approach involves using multiple independent heralded single-photon sources, as used in the experimental verification of spatially overlapping photons for generating entangled states \cite{sun2020experimental,barros2020entangling,Lee:22}. These probabilistic photon sources introduce additional costs in terms of success probability for generating the target state. Alternatively, a deterministic single-photon source, such as a quantum dot, can be employed \cite{Somaschi2016,Schnauber2018,PhysRevLett.89.067901}, which can enhance the success probability. Despite this, implementing deterministic single-photon sources faces practical challenges, including limited spectral tunability, lossy device-to-fiber coupling, and cryogenic cooling requirements \cite{Wang2024}.

In the second step (deformation), one can control and adjust the single photons’ spatial wave functions and polarization (i.e., pseudospin) by using linear optics elements. These include beamsplitters, multimode beamsplitters \cite{Peruzzo2011}, and half-wave plates. These optical components act as deformation operators on the initial state, as described in our theoretical framework.  

In the third step (activation), that is at the detection and verification stage, one can use single-photon counting and quantum state tomography to verify the generated state, as in Refs. \cite{sun2020experimental,barros2020entangling,Lee:22}. As a prospect, it may be valuable to explore the use of quantum witness operators within the sLOCC framework, which could offer a simplified alternative to quantum state tomography, especially in the case of increasingly complex multipartite systems \cite{PhysRevLett.122.120501}.

The goal of this work is to encourage the development of new platforms that can utilize indistinguishability to prepare many-particle entangled states in various experimental configurations beyond free-space optics. For example, superconducting qubits present a promising platform for simulating both bosonic and fermionic systems with two internal and external degrees of freedom. Superconducting ring resonators can function as beam splitters and be used to deform initial states \cite{RevModPhys.93.025005}. Alternatively, qubits in Bose-Einstein and Fermi-Dirac condensates can be prepared at distinct sites within a lattice. Possible internal degrees of freedom in these systems include electrons’ spin or energy levels. Experiments with Bose-Einstein and Fermi-Dirac condensates require suited techniques for single-particle trapping and control. As a deformation operation, particles can be tunneled to other sites similarly to a beam-splitting action. The probability amplitude of this tunneling effect can be adjusted by modifying external parameters such as voltages, magnetic fields, and laser beams \cite{RevModPhys.90.035006, PhysRevLett.118.069901}.}

\section*{Conclusions}

The indistinguishability of identical particles serves as an exploitable resource for generating both bipartite \cite{franco2018indistinguishability,sun2020experimental,barros2020entangling} and multipartite \cite{Lee:22} entangled states. After the sLOCC measurement, the resource state becomes available for various quantum information processing tasks \cite{franco2018indistinguishability, sun2020experimental, castellini2019indistinguishability, Sun2022Activation, bellomo2017n, Activating2019Castellini}, because identical particles become distinguishable and individually addressable by their spatial regions. Based on spatial wave functions, one can devise experimentally feasible entanglement witness operators to test the generated states \cite{PhysRevLett.122.120501}.

{Although maximizing the post-selection probability was not our focus, our scheme for generating multipartite entanglement is more efficient than the heralded-based approach \cite{PhysRevA.102.012604} and the path identity approach \cite{PhysRevLett.118.080401,RevModPhys.94.025007}, both utilizing the indistinguishability of photons. For example, our approach achieved a post-selection probability of $\frac{1}{2^{(N-1)}}$ for the generation of the GHZ state, which is significantly higher than the probability of the path identity approach, given as $\frac{2}{N^{(N/2)}}$ \cite{PhysRevLett.118.080401}. Additionally, we report success probabilities for generating three-partite W states of $\frac{1}{5}$ for bosons and $\frac{1}{3}$ for fermions, improving upon the previously reported probabilities. Specifically, in a linear-optic setup using a three-port splitter, the probability is $\frac{1}{9}$ \cite{Kumar_2023}, and in the no-touching paradigm, it is 0.15 \cite{Blasiak2019}. Also, it is important to note that optimizing the weight matrix adjacency of a given digraph can maximize the post-selection probability. This is the case, for instance, of a bosonic W state \cite{PhysRevA.104.023701} and cluster state \cite{PhysRevA.93.062329}. Noteworthy, this probability of generation W state can further be improved using fermionic statistics, as emerges from our method.}

In this work, we have proposed an algorithmic approach that uses the indistinguishability of identical particles to generate a broad class of multipartite entangled states, including W, Dicke, GHZ, and cluster states. We have represented each generation scheme using bigraphs and digraphs, aligning with the no-label approach for both fermionic and bosonic statistics. This facilitates the examination of the output state, the assessment of its genuine entanglement, and the evaluation of resources in terms of postselection probability. Our approach allows for the search for optimal graph configurations to maximize both the fidelity and postselection probability of the targeted multipartite entangled state. Different graphs may yield the maximum fidelity, such as complete and star digraphs, for the generation of the $W$ state. The graph representation enables the search for arbitrary forms of multipartite state generation by optimizing resources, considering constraints, and using particle statistics. Thus, it serves as a proper interface for optimizing the generation of arbitrary multipartite entangled states with specific probabilities of success and maximum fidelity.

The generation of various multiqubit entangled states via our scheme relies on probabilistic conditions, assuming that a single particle is available upon request. {Also, probabilistic sources can be employed for single-particle preparation in the initial generation stage, such as spontaneous parametric down-conversion (SPDC) and spontaneous four-wave mixing (SFWM) processes in second- and third-order nonlinear materials, respectively \cite{Boyd2023}. However, this source will add an additional cost to sLOCC probability, due to the probabilistic nature of these phenomena.} This is why we explore the trade-off between fidelity and the probability of success, potentially sacrificing the amount of entanglement to achieve a higher sLOCC probability. In prospect, one can also consider experimental imperfections, limitations, and constraints across all three stages and optimize the generation schemes using this algorithmic approach. 

As future perspectives, optimization methods such as simulated annealing could be employed to identify the optimal graph structure with higher success probabilities and fidelity. In conclusion, the exploitability of indistinguishable identical graph nodes in quantum networks offers innovative possibilities within a widely used technology, suggesting further avenues for exploration and study.

\section*{Data Availability}
The main results of this manuscript are analytical and numerical. All data generated or analyzed during this study are included in this article. The code is available upon request.

\section*{Acknowledgments}
R.M. acknowledges support by the Natural Sciences and Engineering Research Council of Canada (NSERC) through the  Discovery and Alliance Grants Schemes, by the MEIE PSR-SIIRI Initiative in Quebec, and by the Canada Research Chair Program.
{R.L.F. acknowledges support by MUR (Ministero dell’Università e della Ricerca) through the following projects: PNRR Project ICON-Q – Partenariato Esteso NQSTI – PE00000023 – Spoke 2, PNRR Project AQuSDIT – Partenariato Esteso SERICS – PE00000014 – Spoke 5, PNRR Project PRISM – Partenariato Esteso RESTART – PE00000001 – Spoke 4}. {F.N. acknowledges support by the I+D+i project MADQuantum-CM, financed by the European Union NextGeneration-EU, Madrid Government and by the PRTR}. {K.M., F.N., and R.L.F. would like to acknowledge Seungbeom Chin and Marcin Karczewski for useful and fruitful discussions}. The authors acknowledge Samira Salehi for her contribution to the illustration of Figure \ref{fig: deformation illustration}.

\appendix
\section{Bell State}\label{App: Bell State}
As mentioned in the main manuscript, spatial deformations alone are insufficient for generating graph states within our formalism. The general deformation alters both the spatial wave function and the pseudospin states of the initial qubits. Here, we discuss the Bell state, represented as:
\begin{equation}\label{sLOCC Bell}
    \ket{\Phi_{+}^{(2)}}=\frac{1}{\sqrt{2}}(\ket{R_{1}\uparrow,R_{2}\uparrow}+\ket{R_{1}\downarrow,R_{2}\downarrow}),
\end{equation}
which requires the general deformation to generate.

First, we consider the initial state in the pure product state form as 
$\ket{\Psi^{(2)}_0}=\ket{\psi_{1}\uparrow}\otimes\ket{\psi_{2}\uparrow}$. The actions of deformations result in each qubit undergoing a transformation: $\ket{\varphi_{1D}}=D_{2}^{(1)}\ket{\psi_{1}\uparrow}=r_{11}\ket{R_{1}\uparrow}+\eta r_{12}\ket{R_{2}\downarrow}$, and $\ket{\varphi_{2D}}=D_{2}^{(2)}\ket{\psi_{2}\uparrow}=r_{21}\ket{R_{1}\downarrow}+r_{22}\ket{R_{2}\uparrow}$  where $r_{ij}$ represent the single-qubit probability amplitudes. As observed in these single-qubit deformed states, the deformation not only alters the spatial wave function but also changes the pseudospins of the initial qubits upon detection in the second region $R_{2}$. The corresponding digraph is similar to Fig.~(2) of the main manuscript, with the difference that the colors of the self-loops differ from the colors of the edges oriented toward them. With probability amplitudes $r_{ij}=\frac{1}{\sqrt{2}}$, the ideal Bell state in \eqref{sLOCC Bell} is generated with an sLOCC probability of $\mathcal{P}{(\ket{\Phi_{+}^{(2)}})}=\frac{1}{2}$.

\section{$W$ state with three and four qubits}\label{appx W state}
In the main part of the paper, we delve into indistinguishability-based generation schemes for the $W$ state with $N$ identical qubits. Here, we provide a detailed explanation of the formalism, illustrating it with examples involving three and four qubits. This exploration allows us to elucidate how the bigraph and digraph picture, employed as a generation scheme, can be utilized to generate the $W$ state.

\subsection{Three-qubit $W$ state}
We begin with an initial state of a three-qubit system expressed in the pure product form as $\ket{\Psi^{(3)}_0}= \ket{\psi_1\uparrow}\otimes\ket{\psi_2\downarrow}\otimes\ket{\psi_3\downarrow}$. Subsequently, we apply spatial deformation, placing each qubit in a superposition of three spatial regions, described by $\ket{\varphi_{iD}}=\sum_{j=1}^{M}r_{ij}\ket{R_j\sigma_{i}}$, where $r_{ij}$ denotes tunable single-qubit probability amplitudes. As a result, the deformed state of the three-qubit system can be written as $\ket{\Psi_D^{(3)}}=\frac{1}{\sqrt{\nu}}\ket{\varphi_{1D},\varphi_{2D},\varphi_{3D}}$, with the normalization factor $\nu=\vert\langle \varphi_{1D},\varphi_{2D},\varphi_{3D}\vert \varphi_{1D},\varphi_{2D},\varphi_{3D}\rangle\vert^2$. It's worth noting that the three-qubit deformed state is not generally factorizable in terms of single-qubit states, given as $\ket{\Psi_D^{(3)}}\neq\frac{1}{\sqrt{\nu}}\ket{\varphi_{1D}}\otimes\ket{\varphi_{2D}}\otimes\ket{\varphi_{3D}}$ \cite{compagnoRSA}. Finally, the sLOCC projection measurement $\hat{\Pi}^{(3)}$ \cite{franco2018indistinguishability,nosrati2020robust} is utilized to project the deformed state onto three separated spatial regions where a single qubit can be found. 

\begin{figure}[!t]
    \centering
\includegraphics[width=1.0\linewidth]{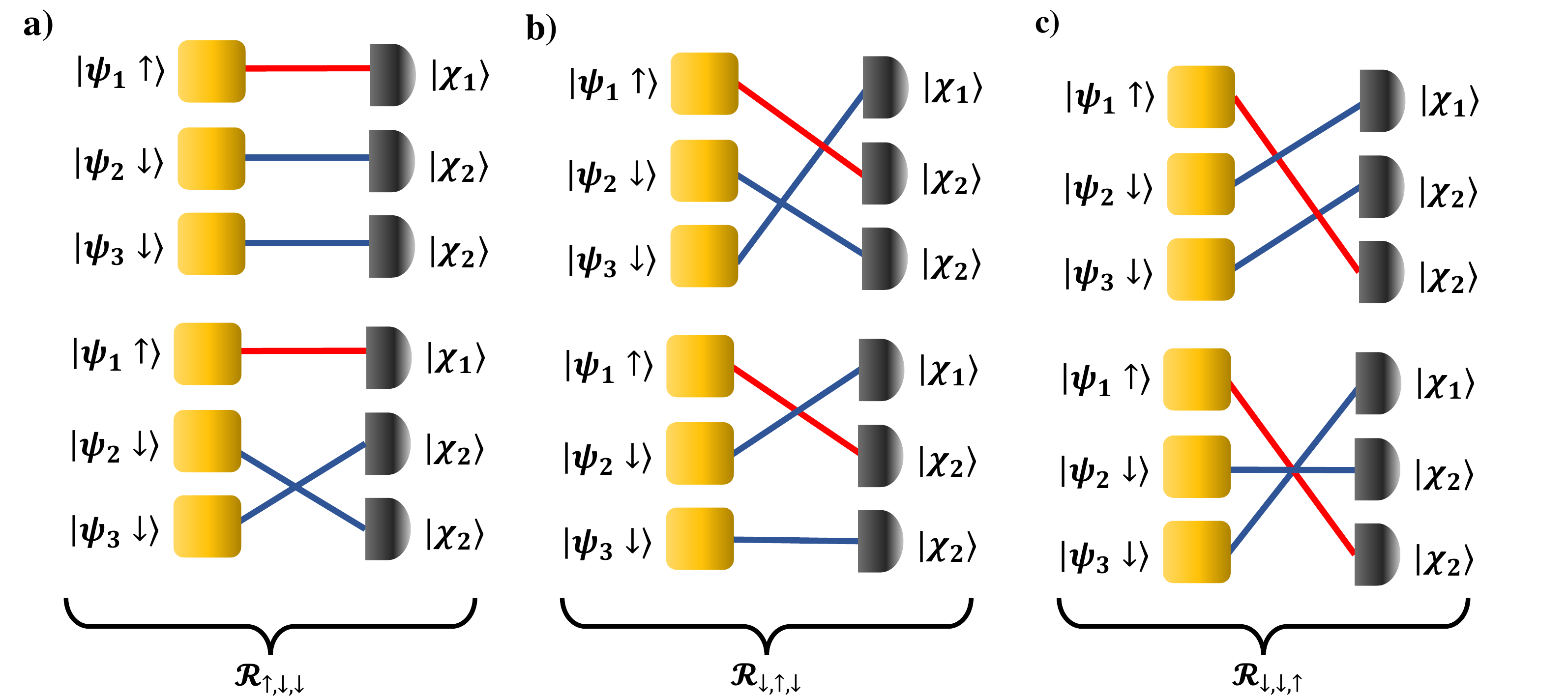}
\caption{Perfect matching sub-bigraphs for generation of the three-particle $W$ state {\cite{Quantumexperimentsandgraphs22017Xuemei,chin2021graph}}. \textbf{a)} Represents the state $\ket{R_1\uparrow,R_2\downarrow,R_3\downarrow}$. \textbf{b)} Represents the state $\ket{R_1\downarrow,R_2\uparrow,R_3\downarrow}$. \textbf{c)} Represents the state $\ket{R_1\downarrow,R_2\downarrow,R_3\uparrow}$.}
    \label{fig: perfect_mactching_W3}
\end{figure}

From a graph theory perspective, we can associate a digraph with the three mentioned generation steps: initialization, deformation, and projection. This assignment is facilitated by the adjacency matrix of a digraph, with elements $\bra{R_i \sigma_i}{\varphi_{jD}}\rangle$, as given by
\begin{equation}
\mathcal{R}_{\sigma_1,\sigma_2,\sigma_3}=
\begin{pmatrix}
    r_{11}\bra{\sigma_1}{\uparrow}\rangle & r_{12}\bra{\sigma_1}{\downarrow}\rangle & r_{13}\bra{\sigma_1}{\downarrow}\rangle\\ r_{21}\bra{\sigma_2}{\uparrow}\rangle & r_{22}\bra{\sigma_2}{\downarrow}\rangle & r_{23}\bra{\sigma_2}{\downarrow}\rangle\\
    r_{31}\bra{\sigma_3}{\uparrow}\rangle & r_{32}\bra{\sigma_3}{\downarrow}\rangle & r_{33}\bra{\sigma_3}{\downarrow}\rangle
\end{pmatrix},    
\end{equation}
which is written over all possible combinations of pseudospins $\sigma_i=\{\uparrow,\downarrow\}$ for $i=1,2,3$. The subsequent steps involve obtaining the determinant of the above adjacency matrix $\left| \mathcal{R}_{\sigma_{1},\sigma_{2},\sigma_{3}}\right|_{\eta}$ and substituting it into the expression $\ket{\Psi_W^{(3)}}=\frac{1}{\sqrt{N_3}}\sum_{\sigma_{1},\dots\sigma_{3}=\{\uparrow,\downarrow\}}\left| \mathcal{R}_{\sigma_{1},\sigma_{2},\sigma_{3}}\right|_{\eta}\ket{R_1\sigma_1,R_2\sigma_2,R_3\sigma_3}$ to generate the desired state $\ket{\Psi_W^{(3)}}$. The explicit expressions of the adjacency matrices with nonzero determinants are given
\begin{equation}
\begin{split}
    &\mathcal{R}_{\uparrow,\downarrow,\downarrow}=
\begin{pmatrix}
    r_{11} & 0 & 0\\ 0 & r_{22} & r_{23}\\
    0 & r_{32} & r_{33}
\end{pmatrix},\quad  \mathcal{R}_{\downarrow,\uparrow,\downarrow}=
\begin{pmatrix}
    0 & r_{12} & r_{13}\\ r_{21} & 0 & 0\\
    0 & r_{32} & r_{33}
\end{pmatrix},
\\
&\mathcal{R}_{\downarrow,\downarrow,\uparrow}=
\begin{pmatrix}
    0 & r_{12} & r_{13}\\ 0 & r_{22} & r_{23}\\
    r_{31} & 0 & 0
\end{pmatrix}. 
\end{split}  
\end{equation}
To generate the state, we connect each of these matrices to perfect matching bigraphs \cite{QuantumExperiments2017Krenn}. For example, the first adjacency matrix corresponds to the scenario where we detect a qubit with an up pseudospin in the first region, while two qubits with down pseudospins are in the second and third regions, or vice versa. The other two adjacency matrices depict the other four perfect matches, as illustrated in Fig.~\ref{fig: perfect_mactching_W3}). Finally, we can write the general form of the three-qubit state as follows:
\begin{equation}\label{eq: General W_3}
\begin{split}
\ket{\Psi_W^{(3)}}=&
\frac{1}{\sqrt{N_3}}\bigg(S_1\ket{R_1\uparrow,R_2\downarrow,R_3\downarrow}+S_2\ket{R_1\downarrow,R_2\uparrow,R_3\downarrow}\\
+&S_3\ket{R_1\downarrow,R_2\downarrow,R_3\uparrow}\bigg),
\end{split}
\end{equation}
where the probability amplitude coefficients $S_1$, $S_2$, and $S_3$ are obtained by determining the determinant of the above adjacency matrices and are specifically given as $|\mathcal{R}_{\uparrow,\downarrow,\downarrow}|_\eta=S_1=r_{11}r_{22}r_{33}+\eta r_{11} r_{23} r_{32}$, $|\mathcal{R}_{\downarrow,\uparrow,\downarrow}|_\eta=S_2=r_{13}r_{21}r_{32}+\eta r_{12}r_{21}r_{33}$ and $|\mathcal{R}_{\downarrow,\downarrow,\uparrow}|_\eta=S_3=r_{12}r_{21}r_{31}+\eta r_{13}r_{22}r_{31}$, with the normalization factor $N_3=\sum_{k=1}^{3}|S_k|^2$.
Now, one can tune the digraph edge weights, $r_{ij}$, to generate the desired state. As discussed in the main manuscript, a specific choice of probability amplitudes is the complete digraph with equal weights, given as $r_{ij}=\frac{1}{\sqrt{3}}$, and bosonic statistics. This, in turn, results in the generation of the three-qubit $W$ state given by $\ket{W^{(3)}}=
\frac{1}{\sqrt{3}}\big(\ket{R_1\uparrow,R_2\downarrow,R_3\downarrow}+\ket{R_1\downarrow,R_2\uparrow,R_3\downarrow}+\:\ket{R_1\downarrow,R_2\downarrow,R_3\uparrow}\big)$, with the sLOCC probability of $\mathcal{P}_{C}\left({\ket{W^{(3)}}}\right)=2/9$. As discussed in the main manuscript, the complete graph is not the only structure that leads to the generation of the $W$ state. Additionally, one can employ the star digraph to generate the $W$ state. The star digraph configuration has an adjacency weight matrix with elements $r_{1i}=\frac{1}{\sqrt{3}}$ (for $i=1,2,3$), $r_{ii}=\frac{1}{\sqrt{2}}$ (for $i=2,3$), and $r_{i1}=\frac{\eta}{\sqrt{2}}$ (for $i=2,3$), applicable to both fermions and bosons. The star digraph configuration leads to sLOCC probabilities   $\mathcal{P}_{star}\left({\ket{W^{(3)}}}\right)=1/5$ for bosons and $\mathcal{P}_{star}\left({\ket{W^{(3)}}}\right)=1/3$ for fermions, respectively. Furthermore, one can adopt the Quantum Fourier Transform (QFT) deformation as well to generate the $W$ state with sLOCC probabilities of   $\mathcal{P}_{QFT}\left({\ket{W^{(3)}}}\right)=1/9$ for bosons and $\mathcal{P}_{QFT}\left({\ket{W^{(3)}}}\right)=1/3$ for fermions. 
\subsection{Four-qubit $W$ state}

The next step involves considering a four-qubit system, which leads to diverse configurations in generations of quantum states. Similarly, we assume the initial state of a four-qubit in the pure product state as $\ket{\Psi^{(4)}_0} = \ket{\psi_1\uparrow}\otimes\ket{\psi_2\downarrow}\otimes\ket{\psi_3\downarrow}\otimes\ket{\psi_4\downarrow}$. Similar to the three-qubit case, we aim to come up with explicit expression of the adjacency matrices $\mathcal{R}_{\sigma_1,\sigma_2,\sigma_3,\sigma_4}$. The adjacency matrices, connected to perfect matching bigraphs with nonzero determinants are
\begin{equation}
\begin{split}
\mathcal{R}_{\uparrow,\downarrow,\downarrow,\downarrow}=\begin{pmatrix}
r_{11} & 0 & 0 & 0 \\
0 & r_{22} & r_{23} & r_{24} \\
0 & r_{32} & r_{33} & r_{34} \\
0 & r_{42} & r_{43} & r_{44} \\
\end{pmatrix},\:\: \mathcal{R}_{\downarrow,\uparrow,\downarrow,\downarrow}=\begin{pmatrix}
0 & r_{12} & r_{13} & r_{14} \\
r_{21} & 0 & 0 & 0 \\
0 & r_{32} & r_{33} & r_{34} \\
0 & r_{42} & r_{43} & r_{44} \\
\end{pmatrix},
\\
\mathcal{R}_{\downarrow,\downarrow,\uparrow,\downarrow}=\begin{pmatrix}
0 & r_{12} & r_{13} & r_{14} \\
0 & r_{22} & r_{23} & r_{24} \\
r_{31} & 0 & 0 & 0 \\
0 & r_{42} & r_{43} & r_{44} \\
\end{pmatrix},\:\:\mathcal{R}_{\downarrow,\downarrow,\downarrow,\uparrow}=\begin{pmatrix}
0 & r_{12} & r_{13} & r_{14} \\
0 & r_{22} & r_{23} & r_{24} \\
0 & r_{32} & r_{33} & r_{34} \\
r_{41} & 0 & 0 & 0 \\
\end{pmatrix}.
\end{split}
\end{equation}
After obtaining the determinant-like of the above adjacency matrices, we can the generated four-qubit state as
\begin{equation}\label{eq: W4_general}
\begin{split}
    \ket{\Psi_W^{(4)}}=
    \frac{1}{\sqrt{N_4}} \bigg(&S_{1}\ket{R_{1}\uparrow,R_{2}\downarrow,R_{3}\downarrow,R_{4}\downarrow}
    \\
    +&S_2\ket{R_{1}\downarrow,R_{2}\uparrow,R_{3}\downarrow,R_{4}\downarrow}
    \\
+&S_{3}\ket{R_{1}\downarrow,R_{2}\downarrow,R_{3}\uparrow,R_{4}\downarrow}
\\+&S_{4}\ket{R_{1}\downarrow,R_{2}\downarrow,R_{3}\downarrow,R_{4}\uparrow}
    \bigg),
\end{split}
\end{equation}
where probability amplitude coefficients are $S_1=|\mathcal{R}_{\uparrow,\downarrow,\downarrow,\downarrow}|_\eta$, $S_2=|\mathcal{R}_{\downarrow,\uparrow,\downarrow,\downarrow}|_\eta$, $S_3=|\mathcal{R}_{\downarrow,\downarrow,\uparrow,\downarrow}|_\eta$, and $S_4=|\mathcal{R}_{\downarrow,\downarrow,\downarrow,\uparrow}|_\eta$ with the normalization factor as $N_4=\sum_i^4|S_i|^2$.
Similarly, we can generate the four-qubit $W$-state $\ket{W^{(4)}}=
\frac{1}{2}\big(\ket{R_{1}\uparrow,R_2\downarrow,R_3\downarrow,R_4\downarrow}+\ket{R_1\downarrow,R_2\uparrow,R_3\downarrow,R_4\downarrow}+\ket{R_1\downarrow,R_2\downarrow,R_3\uparrow,R_4\downarrow}+\ket{R_1\downarrow,R_2\downarrow,R_3\downarrow,R_4\uparrow}\big)$ using the complete (equal weights), star, and QFT digraph configurations with sLOCC probabilities of $\mathcal{P}_C\left({\ket{W^{(4)}}}\right)=3/32$ (for bosons), $\mathcal{P}_{star}\left({\ket{W^{(4)}}}\right)=1/16$ (for bosons), $\mathcal{P}_{star}\left({\ket{W^{(4)}}}\right)=1/4$ (for fermions), and $\mathcal{P}_{QFT}\left({\ket{W^{(4)}}}\right)=1/4$ (for fermions). However, the state generated with QFT deformation for bosons is expressed as follows
\begin{equation}
    \begin{split}
\ket{\Psi_{W-QFT}^{(4)}}=
\frac{1}{2}\bigg(&\ket{R_{1}\uparrow,R_2\downarrow,R_3\downarrow,R_4\downarrow}-\\
&\ket{R_1\downarrow,R_2\uparrow,R_3\downarrow,R_4\downarrow}-
\\
&\ket{R_1\downarrow,R_2\downarrow,R_3\uparrow,R_4\downarrow}+\\
&\ket{R_1\downarrow,R_2\downarrow,R_3\downarrow,R_4\uparrow}\bigg).
\end{split}
\end{equation}
Although the state above is entangled among the four qubits, it does not have the form of the $W$ state. 
\begin{table*}
\normalsize
  \label{tab:$W$ state}
  \begin{tabular}
      {|c|c|c|c|c|} \hline Graphs & Fidelity for bosons  & Fidelity for fermions & Probability for bosons & Probability for fermions 
      \\\hline
      \parbox[c]{1.4in}{
      \includegraphics[width=1.2in]{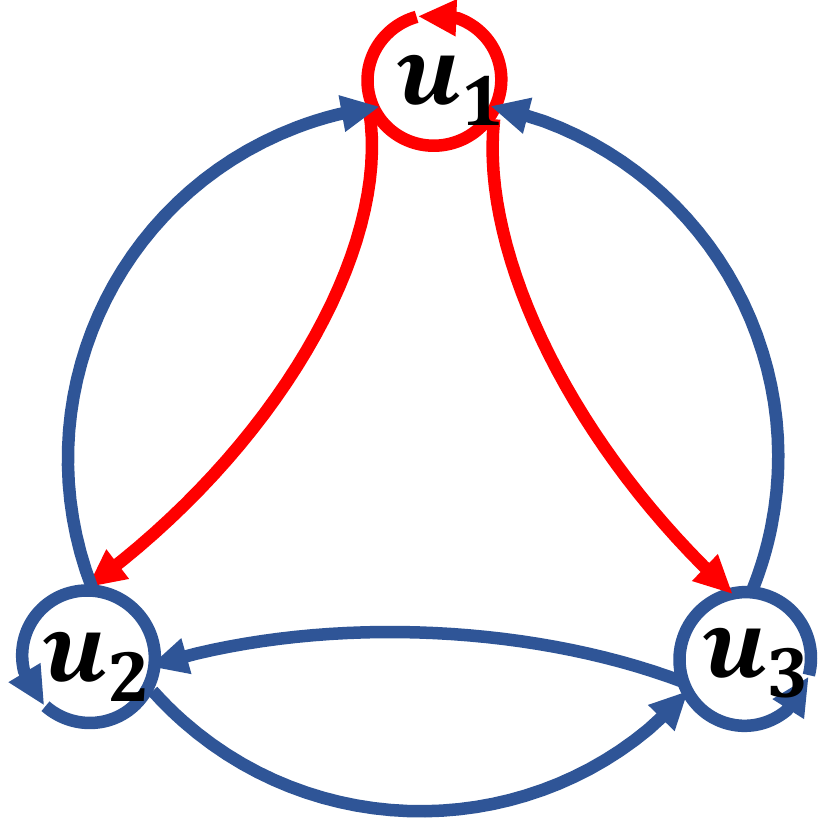}} & $F\left({\ket{W^{(3)}}}\right)=1$ & $F\left({\ket{W^{(3)}}}\right)=0$ & $\mathcal{P}_{C}\left({\ket{W^{(3)}}}\right)=2/9$ & $\mathcal{P}_{C}\left({\ket{W^{(3)}}}\right)=0$ \\\hline
      \parbox[c]{1.2in}{
      \includegraphics[width=1.2in]{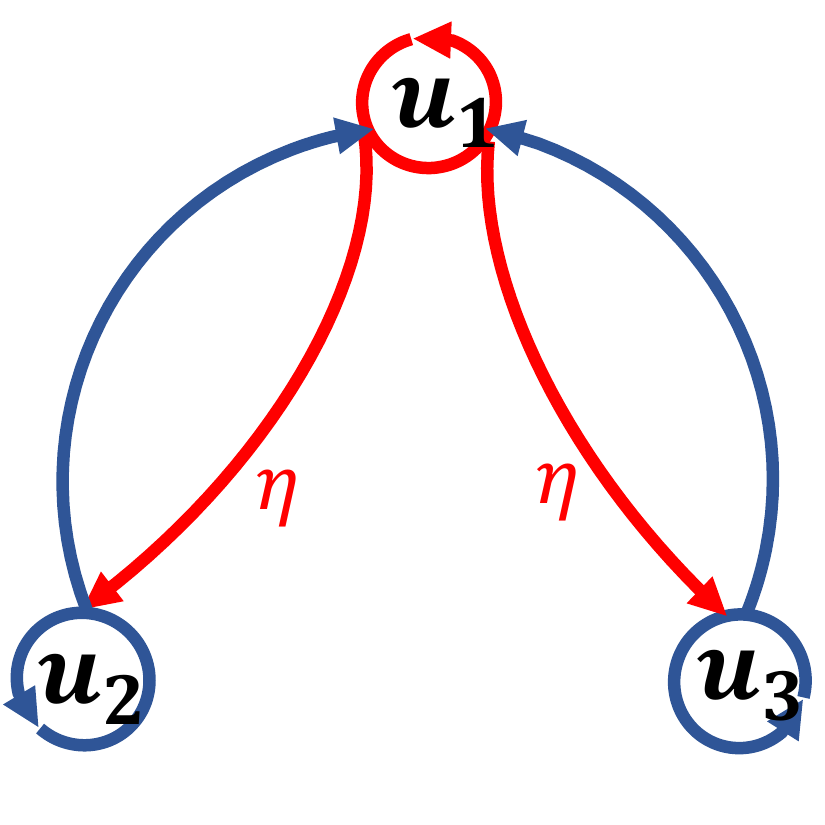}} & $F\left({\ket{W^{(3)}}}\right)=1$ & $F\left({\ket{W^{(3)}}}\right)=1$ & $\mathcal{P}_{star}\left({\ket{W^{(3)}}}\right)=1/5$ & $\mathcal{P}_{star}\left({\ket{W^{(3)}}}\right)=1/3$ 
      \\\hline
      \parbox[c]{1.2in}{
      \includegraphics[width=1.2in]{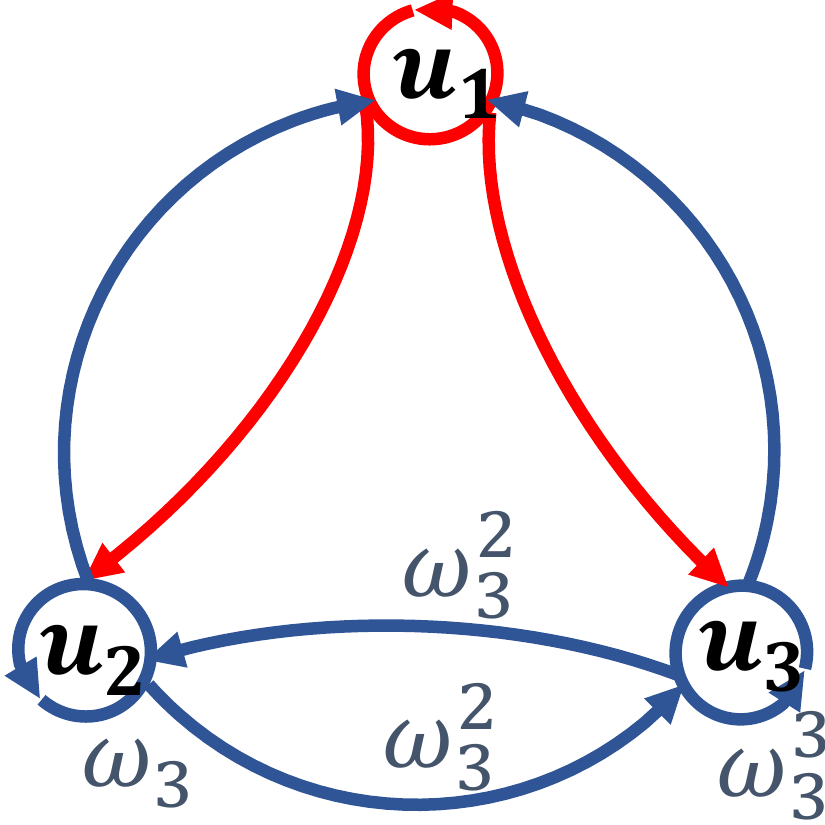}} & $F\left({\ket{W^{(3)}}}\right)=1$ & $F\left({\ket{W^{(3)}}}\right)=1$ & $\mathcal{P}_{QFT}\left({\ket{W^{(3)}}}\right)=1/9$ & $\mathcal{P}_{QFT}\left({\ket{W^{(3)}}}\right)=1/3$ 
      \\\hline
      \parbox[c]{1.2in}{
      \includegraphics[width=1.2in]{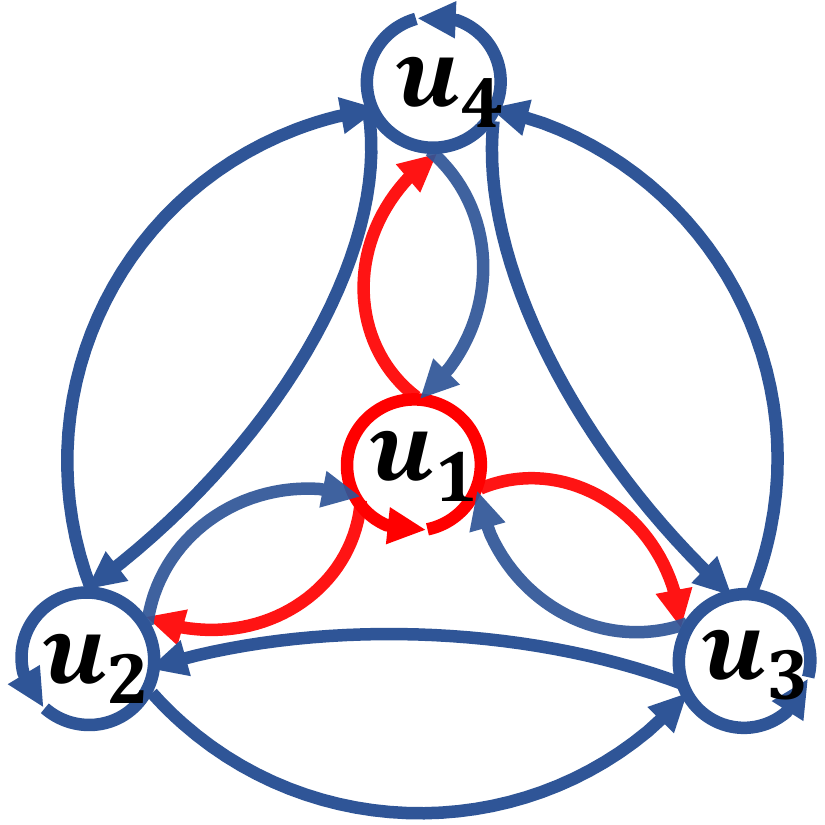}} & $F\left({\ket{W^{(4)}}}\right)=1$ & $F\left({\ket{W^{(4)}}}\right)=1$ & $\mathcal{P}_{C}\left({\ket{W^{(4)}}}\right)=3/32$ & $\mathcal{P}_{C}\left({\ket{W^{(4)}}}\right)=0$ \\\hline
      \parbox[c]{1.2in}{
      \includegraphics[width=1.2in]{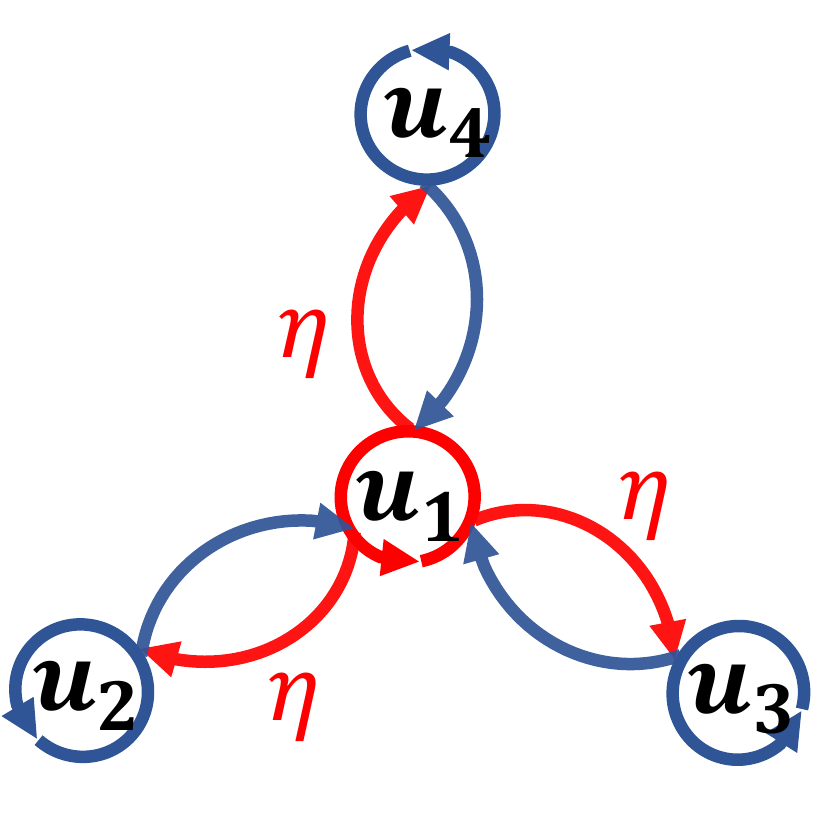}} & $F\left({\ket{W^{(4)}}}\right)=1$ & $F\left({\ket{W^{(4)}}}\right)=1$ & $\mathcal{P}_{star}\left({\ket{W^{(4)}}}\right)=1/16$ & $\mathcal{P}_{star}\left({\ket{W^{(4)}}}\right)=1/4$ 
      \\\hline
      \parbox[c]{1.2in}{
      \includegraphics[width=1.2in]{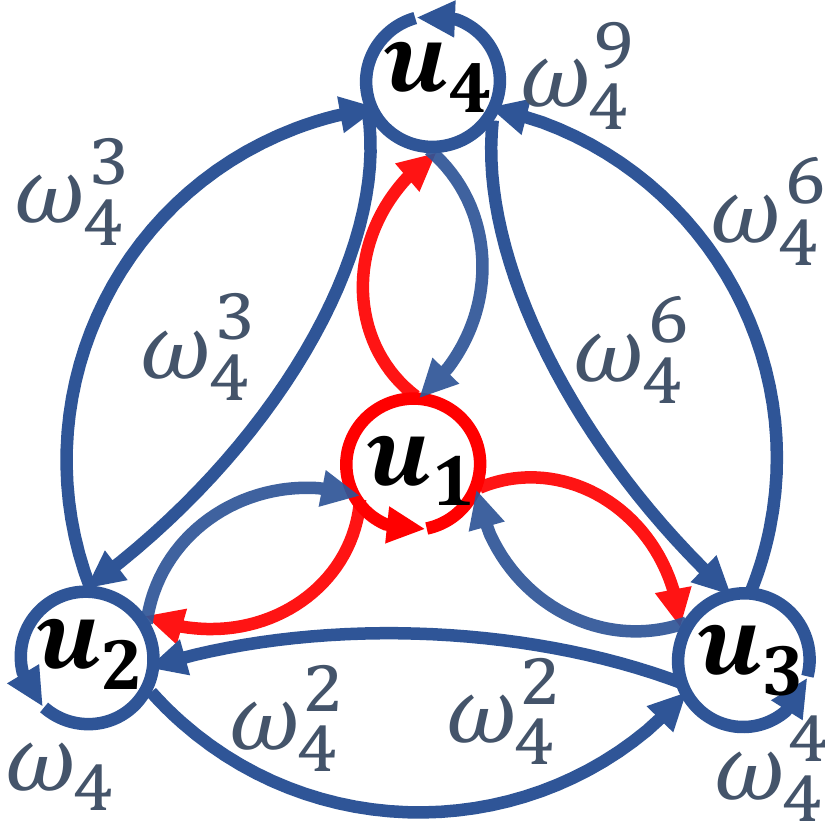}} & $F\left({\ket{W^{(4)}}}\right)=0$ & $F\left({\ket{W^{(4)}}}\right)=1$ & $\mathcal{P}_{QFT}\left({\ket{W^{(4)}}}\right)=1/16$ & $\mathcal{P}_{QFT}\left({\ket{W^{(4)}}}\right)=1/4$
      \\\hline
      \parbox[c]{1.2in}{
      \includegraphics[width=1.2in]{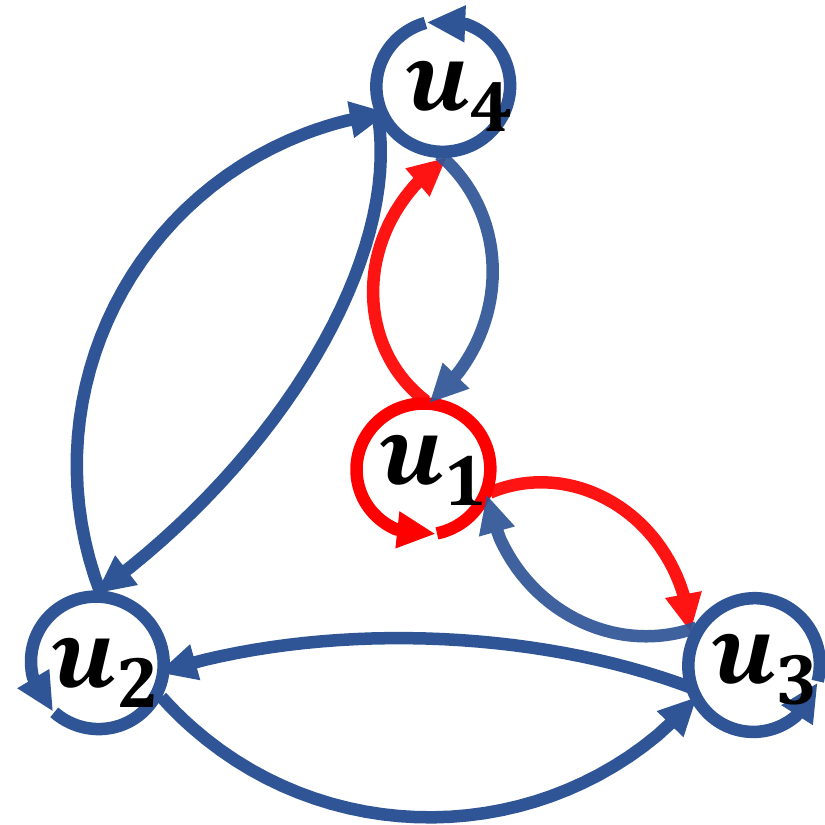}} & $F\left({\ket{W^{(4)}}}\right)=3/4$ & $F\left({\ket{W^{(4)}}}\right)=3/4$ & $\mathcal{P}_{chain}\left({\ket{W^{(4)}}}\right)=0.1139$ & $\mathcal{P}_{chain}\left({\ket{W^{(4)}}}\right)=0.1429$
      \\\hline
  \end{tabular}
  \caption{\textbf{Digraph configurations for generating $W$ states with three and four qubits.} This table outlines diverse digraph configurations utilized to generate four-qubit $W$ states involving identical bosonic and fermionic qubits. It includes their corresponding fidelities ($F$) and sLOCC probabilities ($\mathcal{P}$). The presence of $\eta$, highlighted in red, shows a $\pi$ phase shift between interconnected nodes for fermions. Moreover, edge weights labeled with $\omega_N$ denote $e^{2i\pi/N}$.} 
\end{table*}
One can explore alternative graph configurations, such as a closed-chained digraph. In this configuration, each node has three edges, including a self-loop. In this scenario, each qubit overlaps equally with its nearest neighbor nodes. Specifically, the qubit with a down pseudospins undergoes spatial deformation and takes the single-qubit state form $\ket{\varphi_{1D}}=\frac{1}{\sqrt{3}}\left(\ket{R_1\uparrow}+\eta\ket{R_3\uparrow}+\eta\ket{R_4\uparrow}\right)$. The other qubits with an up pseudospins undergo deformation and take the form $\ket{\varphi_{2D}}=\frac{1}{\sqrt{3}}\left(\ket{R_{2}\downarrow}+\ket{R_{3}\downarrow}+\ket{R_{4}\downarrow}\right)$, $\ket{\varphi_{3D}}=\frac{1}{\sqrt{3}}\left(\ket{R_{3}\downarrow}+\ket{R_{2}\downarrow}+\ket{R_{1}\downarrow}\right)$, and $\ket{\varphi_{4D}}=\frac{1}{\sqrt{3}}\left(\ket{R_{4}\downarrow}+\ket{R_{2}\downarrow}+\ket{R_{1}\downarrow}\right)$. Within the choice of these $r_{ij}$, the generated state with closed-chain configuration takes the following form:
\begin{equation}
    \begin{split}
\ket{\Psi_{W-chain}^{(4)}}=&
\frac{1}{\sqrt{3}}\bigg(\ket{R_{1}\uparrow,R_2\downarrow,R_4\downarrow}+\ket{R_1\downarrow,R_2\uparrow,\downarrow,R_4\downarrow}
\\
+&\ket{R_1\downarrow,R_2\downarrow,R_4\uparrow}\bigg)\otimes\ket{R_3\downarrow},
\end{split}
\end{equation}
The state mentioned above is a biproduct of a three-qubit state, in the form of a $W$ state, and a single-qubit state. This indicates that only three qubits are entangled with a fidelity of $F\left({\ket{W^{(4)}}}\right) = 3/4$, which does not violate the threshold of $\frac{N-1}{N}$ for four qubits.  In summary, we present the various generation configuration schemes as digraphs with their corresponding fidelity and success (sLOCC) probabilities for both bosonic and fermionic $W$ states involving three and four qubits in Table~\ref{tab:$W$ state}. 
\section{Symmetric Dicke state of four particles}\label{appendix: Dicke state}
\begin{table*}
\normalsize
  \label{tab:Dicke state}
  \begin{tabular}
      {|c|c|c|c|c|} \hline Graphs & Fidelity for bosons  & Fidelity for fermions & Probability for bosons & Probability for fermions 
      \\\hline
      \parbox[c]{1.2in}{
      \includegraphics[width=1.2in]{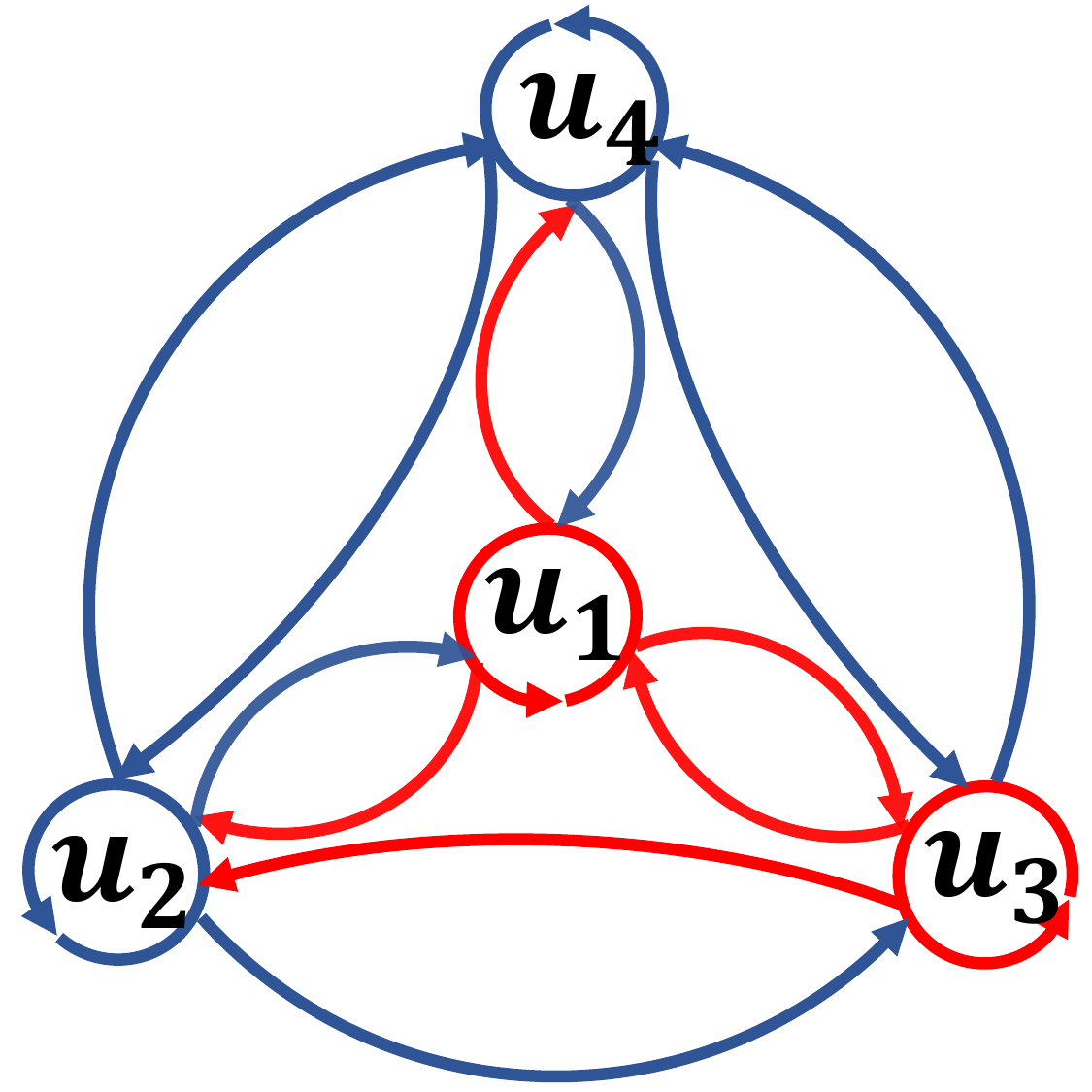}} & $F\left({\ket{D^{(4)}}}\right)=1$ & $F\left({\ket{D^{(4)}}}\right)=1$ & $\mathcal{P}_{C}\left({\ket{D^{(4)}}}\right)=0.0938$ & $\mathcal{P}_{C}\left({\ket{D^{(4)}}}\right)=0$ \\\hline
      \parbox[c]{1.2in}{
      \includegraphics[width=1.2in]{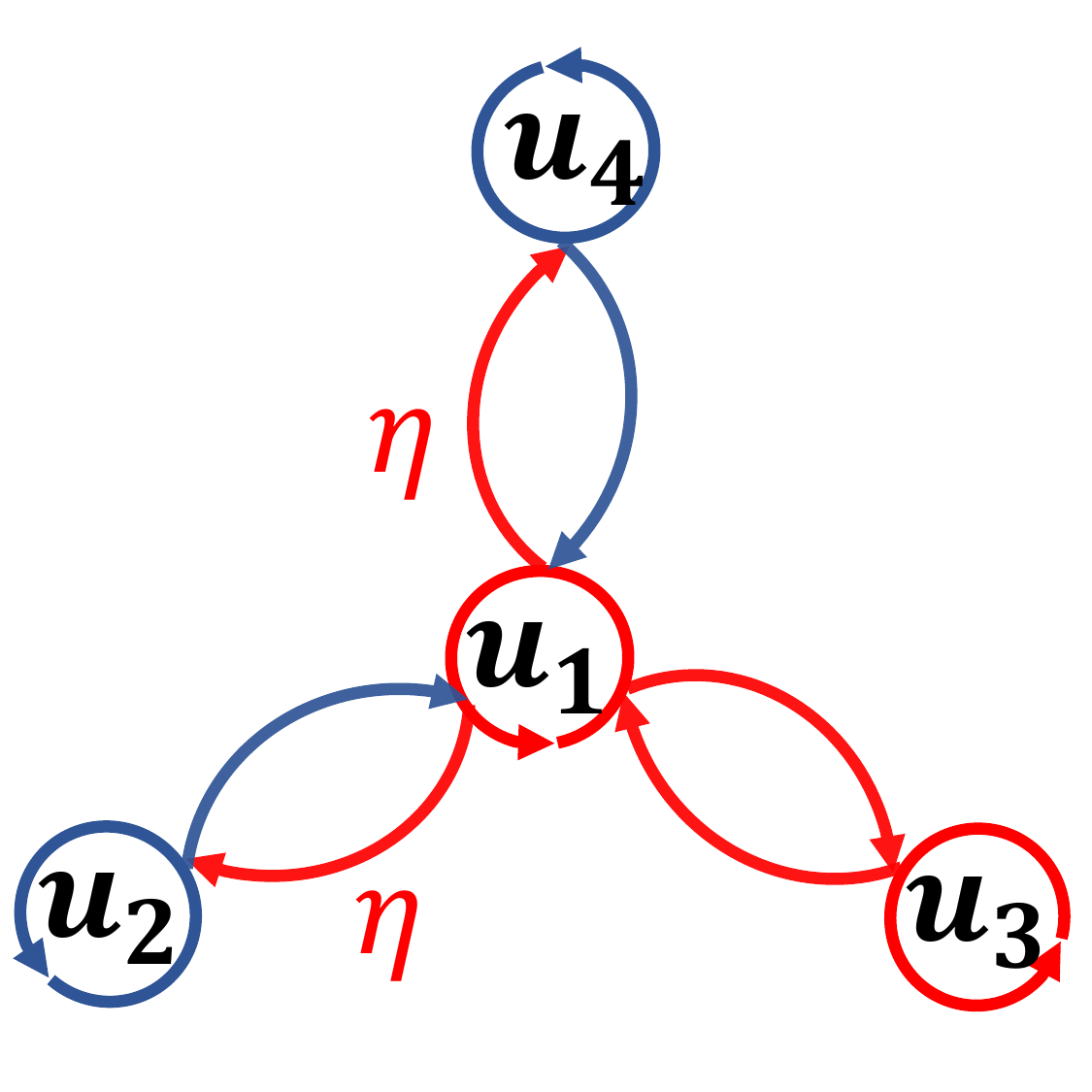}} & $F\left({\ket{D^{(4)}}}\right)=4/9$ & $F\left({\ket{D^{(4)}}}\right)=4/9$ & $\mathcal{P}_{star}\left({\ket{W^{(4)}}}\right)=1/10$ & $\mathcal{P}_{star}\left({\ket{D^{(4)}}}\right)=1/4$ 
      \\\hline
      \parbox[c]{1.2in}{
      \includegraphics[width=1.2in]{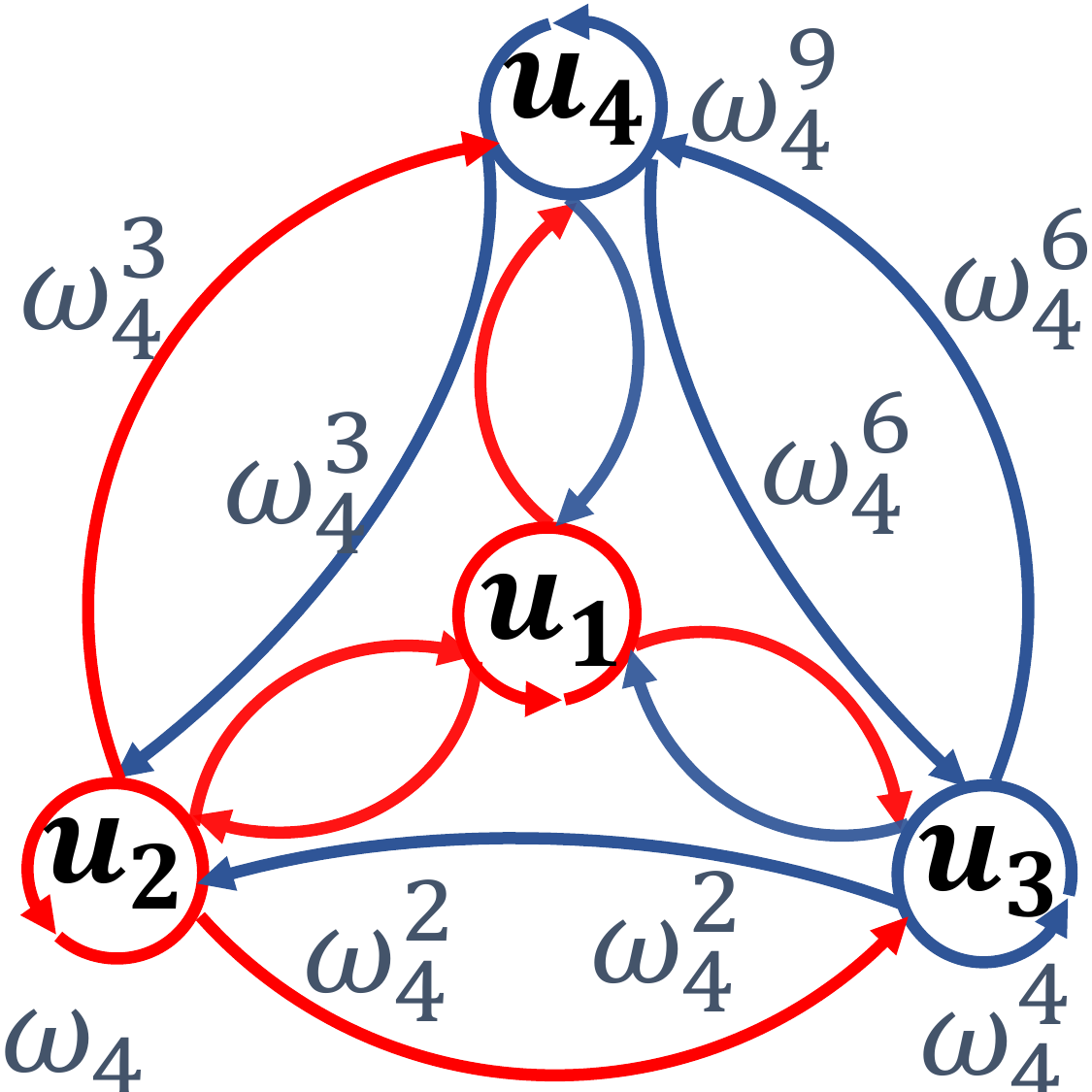}} & $F\left({\ket{D^{(4)}}}\right)=0$ & $F\left({\ket{D^{(4)}}}\right)=2/3$ & $\mathcal{P}_{QFT}\left({\ket{D^{(4)}}}\right)=0.1250$ & $\mathcal{P}_{QFT}\left({\ket{D^{(4)}}}\right)=1/4$
      \\\hline
      \parbox[c]{1.2in}{
      \includegraphics[width=1.2in]{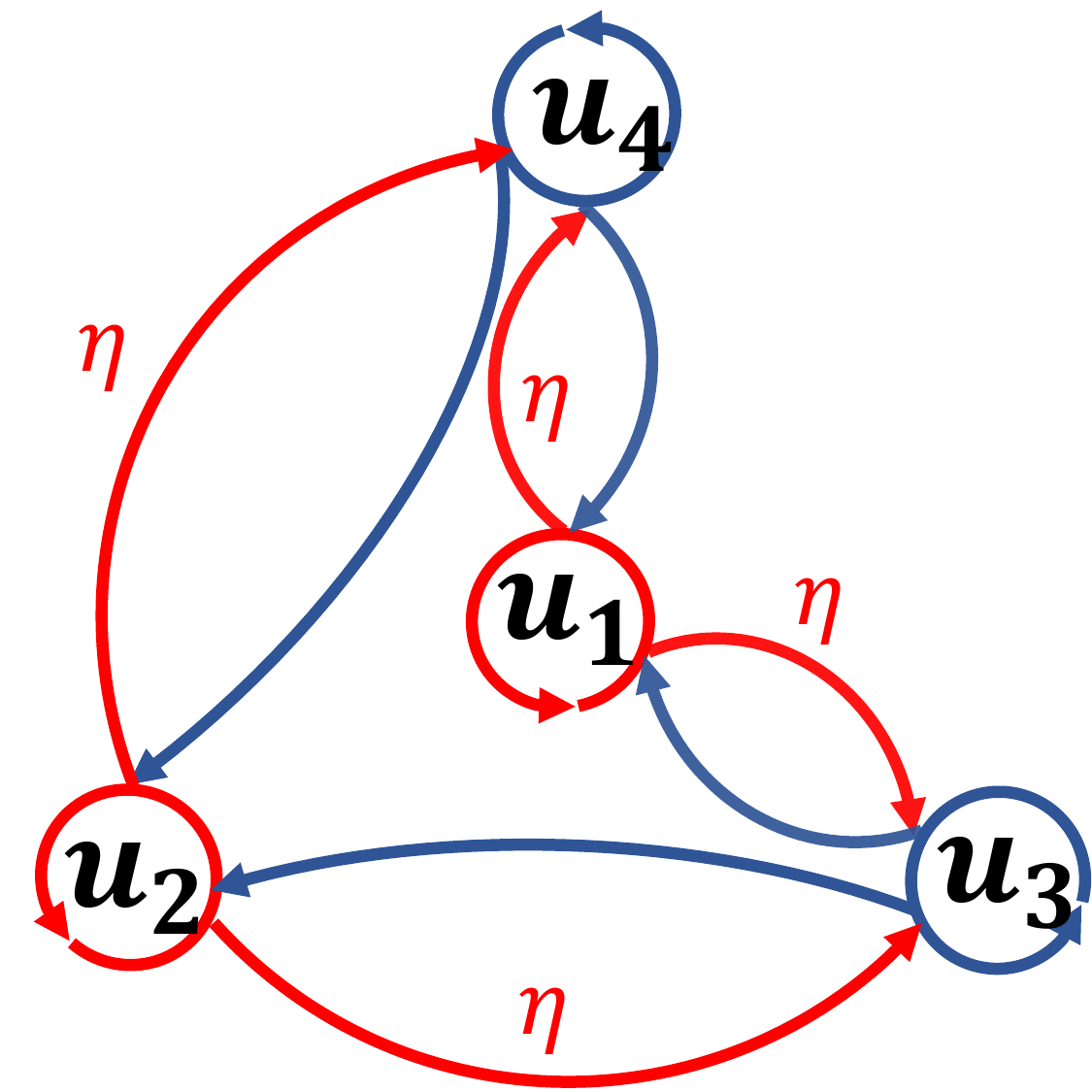}} & $F\left({\ket{D^{(4)}}}\right)=0.6429$ & $F\left({\ket{D^{(4)}}}\right)=3/4$ & $\mathcal{P}_{chain}\left({\ket{D^{(4)}}}\right)=0.1234$ & $\mathcal{P}_{chain}\left({\ket{D^{(4)}}}\right)=0.1429$
      \\\hline
  \end{tabular}
  \caption{\textbf{Digraph configurations for generating four-qubit Dicke states.} The table summarizes digraph configurations utilized to generate four-qubit Dicke states, involving identical bosonic and fermionic qubits, along with their corresponding fidelities ($F$) and sLOCC probabilities ($\mathcal{P}$). The presence of $\eta$, highlighted in red, indicates a $\pi$ phase shift between interconnected nodes for fermions. Moreover, edge weights labeled with $\omega_N$ denote $e^{2i\pi/N}$.} 
\end{table*}

We initially consider four identical qubits in product state form: $\ket{\Psi^{(4)}}=\ket{\psi_1\uparrow}\otimes\ket{\psi_2\downarrow}\otimes\ket{\psi_3\uparrow}\otimes\ket{\psi_4\downarrow}$. We can associate the adjacency matrix of a digraph with the three-generation steps: initialization, deformation, and projection, as follows:
\begin{equation}\label{eq: 4-qubit D matrix}
\mathcal{R}_{\sigma_1,\dots,\sigma_4}=
\begin{pmatrix}
    r_{11}\bra{\sigma_1}{\uparrow}\rangle & r_{12}\bra{\sigma_1}{\downarrow}\rangle & r_{13}\bra{\sigma_1}{\uparrow}\rangle & r_{14}\bra{\sigma_1}{\downarrow}\rangle \\ r_{21}\bra{\sigma_2}{\uparrow}\rangle & r_{22}\bra{\sigma_2}{\downarrow}\rangle & r_{23}\bra{\sigma_2}{\uparrow}\rangle & r_{24}\bra{\sigma_2}{\downarrow}\rangle\\
    r_{31}\bra{\sigma_3}{\uparrow}\rangle & r_{32}\bra{\sigma_3}{\downarrow}\rangle & r_{33}\bra{\sigma_3}{\uparrow}\rangle & 
    r_{34}\bra{\sigma_3}{\downarrow}\rangle \\
    r_{41}\bra{\sigma_4}{\uparrow}\rangle & r_{42}\bra{\sigma_4}{\downarrow}\rangle & r_{43}\bra{\sigma_4}{\uparrow}\rangle & r_{44}\bra{\sigma_4}{\downarrow}\rangle
\end{pmatrix},    
\end{equation}
resulting in the general state of $\ket{\Psi_{D}^{(4)}}=\frac{1}{\sqrt{N_4}}\sum_{\sigma_{1},\dots\sigma_{4}=\{\uparrow,\downarrow\}}\left| \mathcal{R}_{\sigma_{1},\dots,\sigma_{4}}\right|_{\eta}\ket{R_1\sigma_1,R_2\sigma_2,R_3\sigma_3,R_4\sigma_4}$ with $N_4=\sum_{\sigma_{1},\dots\sigma_{4}=\{\uparrow,\downarrow\}}\left| \mathcal{R}_{\sigma_{1},\dots,\sigma_{4}}\right|_{\eta}$ as a normalization factor. The non-zero determinants occur with six different combinations of pseudospin $\{\sigma_{1},\dots\sigma_{4}\}=\{\uparrow,\downarrow\}$, which result in the general form of the generated state, given in the following expression:
\begin{equation}\label{eq: D4_general}
    \begin{split}
        \ket{\Psi_{D}^{(4)}}=&\frac{1}{\sqrt{N_4}}\bigg(S_1\ket{R_1\uparrow,R_2\downarrow,R_3\uparrow,R_4\downarrow}
        \\
+&S_2\ket{R_1\uparrow,R_2\uparrow,R_3\downarrow,R_4\downarrow}+S_3\ket{R_1\uparrow,R_2\downarrow,R_3\downarrow,R_4\uparrow}
        \\
+&S_4\ket{R_1\downarrow,R_2\uparrow,R_3\downarrow,R_4\uparrow}+S_5\ket{R_1\downarrow,R_2\uparrow,R_3\uparrow,R_4\downarrow} 
\\
+&S_6\ket{R_1\downarrow,R_2\downarrow,R_3\uparrow,R_4\uparrow}\bigg),
    \end{split}
\end{equation}
where probability amplitudes are given by the determinants of digraph adjacency matrices as $S_1=|\mathcal{R}_{\uparrow,\downarrow,\uparrow,\downarrow,}|_\eta$, $S_2=|\mathcal{R}_{\uparrow,\uparrow,\downarrow,\downarrow}|_\eta$, $S_3=|\mathcal{R}_{\uparrow,\downarrow,\downarrow,\uparrow}|_\eta$, and $S_4=|\mathcal{R}_{\downarrow,\uparrow,\downarrow,\uparrow}|_\eta$, $S_5=|\mathcal{R}_{\downarrow,\uparrow,\uparrow,\downarrow}|_\eta$, and $S_6=|\mathcal{R}_{\downarrow,\downarrow,\uparrow,\uparrow}|_\eta$. 

As mentioned in the main part of the manuscript, the complete digraph is when qubits are distributed over all spatial regions with equal probability amplitudes of $r_{ij}=\frac{1}{2}$ ($i,j=1,2,3,4$). With such a generation setting, it leads to a symmetric Dicke state with a sLOCC probability of $\mathcal{P}_{C}\left({\ket{D^{(4)}}}\right)=0.0938$ for bosons. However, this configuration does not lead to any Dicke state for fermions. 

Alternatively, one may explore the on-site design utilizing the QFT operator to generate the symmetric Dicke state with four qubits. This involves defining $r_{ij}=e^{\mathbf{i}(i-1)(j-1)\frac{\pi}{2}}$ in the matrix adjacency as shown in Eq. \ref{eq: 4-qubit D matrix}. With this choice, the resulting state is expressed as:
\begin{equation}
    \begin{split}
    \ket{\Psi_{D-QFT}^{(4)}}=\frac{1}{2}\Big(&\ket{R_1\uparrow,R_2\downarrow,R_3\downarrow,R_4\uparrow}+
    \\&\ket{R_1\downarrow,R_2\uparrow,R_3\uparrow,R_4\downarrow}-
    \\
    &\ket{R_1\uparrow,R_2\uparrow,R_3\downarrow,R_4\downarrow}-\\
    &\ket{R_1\downarrow,R_2\downarrow,R_3\uparrow,R_4\uparrow}\Big).
    \end{split}
\end{equation}
As can be seen, the state above is missing the terms $\ket{R_1\uparrow,R_2\downarrow,R_3\uparrow,R_4\downarrow}$ and $\ket{R_1\downarrow,R_2\uparrow,R_3\downarrow,R_4\uparrow}$, thus not matching the symmetric Dicke state.

Also, let us consider the star digraph configuration as another example. With the star digraph, the generated state takes the following form:
\begin{equation}
\begin{split}
\ket{\Psi_{D-star}^{(4)}}=&
\frac{1}{\sqrt{6}}\bigg(2\ket{R_{1}\uparrow,R_2\downarrow,R_4\downarrow}+\ket{R_1\downarrow,R_2\uparrow,\downarrow,R_4\downarrow}
\\
+&\ket{R_1\downarrow,R_2\downarrow,R_4\uparrow}\bigg)\otimes\ket{R_3\uparrow},
\end{split}
\end{equation}
with a sLOCC probability of  $\mathcal{P}_{star}\left({\ket{D^{(4)}}}\right)=0.1$ for bosons and $\mathcal{P}_{star}\left({\ket{D^{(4)}}}\right)=0.25$ for fermions. However, as observed, the state is not a four-partite entangled Dicke state. 

Furthermore, we can consider a closed chain configuration in digraph, which results in the following states for bosons and fermions, respectively:
\begin{equation}\label{eq: D4_C}
	\begin{split}
	\ket{\Psi_{D-chain}^{(4)}}=&
		\frac{1}{\sqrt{21}}\bigg(4\ket{R_1\uparrow,R_2\downarrow,R_3\uparrow,R_4\downarrow}
  \\
+&\ket{R_1\uparrow,R_2\uparrow,R_3\downarrow,R_4\downarrow}+\ket{R_1\uparrow,R_2\downarrow,R_3\downarrow,R_4\uparrow}
  \\ +&\ket{R_1\downarrow,R_2\uparrow,R_3\downarrow,R_4\uparrow}+\ket{R_1\downarrow,R_2\uparrow,R_3\uparrow,R_4\downarrow}
  \\+&\ket{R_1\downarrow,R_2\downarrow,R_3\uparrow,R_4\uparrow}\bigg),
	\end{split}
\end{equation}
\begin{equation}
    \begin{split}
\ket{\Psi_{D-chain}^{(4)}}=&\frac{1}{\sqrt{5}}\bigg(\ket{R_1\uparrow,R_2\downarrow,R_3\uparrow,R_4\downarrow}
\\
+&\ket{R_1\uparrow,R_2\uparrow,R_3\downarrow,R_4\downarrow}+\ket{R_1\uparrow,R_2\downarrow,R_3\downarrow,R_4\uparrow}
\\
+&\ket{R_1\downarrow,R_2\uparrow,R_3\uparrow,R_4\downarrow}
\\
+&\ket{R_1\downarrow,R_2\downarrow,R_3\uparrow,R_4\uparrow}\bigg).
    \end{split}
\end{equation}
Here, bosonic and fermionic states have fidelity values of $F\left({\ket{D^{(4)}}}\right)=0.6429$ and $F\left({\ket{D^{(4)}}}\right)=3/4$ and sLOCC probabilities of $\mathcal{P}_{D-chain}\left({\ket{D^{(4)}}}\right)=0.1243$ and  $\mathcal{P}_{D-chain}\left({\ket{D^{(4)}}}\right)=0.1429$, respectively. Finally, we summarize all generation settings and their associated fidelities and sLOCC probabilities for both bosons and fermions in Table~\ref{tab:Dicke state}.



%

\end{document}